\documentclass[11pt]{article}
\usepackage{amssymb}
\usepackage[numbers,sort&compress]{natbib}
\usepackage{amsmath,amsfonts,graphicx,epsfig}
\usepackage{bm}
\usepackage[dvipsnames]{xcolor}
\definecolor{lgray}{gray}{0.35}
\usepackage[colorlinks=true,urlcolor=Gray,linkcolor=lgray,
citecolor=Gray,
             pdfpagelabels=true,hypertexnames=true,
            plainpages=false,naturalnames=false,
             ]{hyperref}
\newcommand{\be}{\begin{equation}}
\newcommand{\ee}{\end{equation}}
\newcommand{\bea}{\begin{eqnarray}}
\newcommand{\eea}{\end{eqnarray}}
\newcommand{\nn}{\nonumber}

\newcommand{\x}{{\boldsymbol x}}
\newcommand{\y}{{\boldsymbol y}}
\newcommand{\z}{{\boldsymbol z}}
\newcommand{\w}{{\boldsymbol w}}
\newcommand{\K}{{\boldsymbol k}}

\newcommand{\q}{{\boldsymbol q}}
\newcommand{\n}{ \hat{n} }
\newcommand{\F}{{\cal F}}
\newcommand{\de}{{\rm d}}
\newcommand{\PDF}{{\rm P}}

\begin{document}

\setlength\arraycolsep{2pt}

\renewcommand{\theequation}{\arabic{section}.\arabic{equation}}
\setcounter{page}{1}

\begin{titlepage}

\begin{center}

{\huge \bf Non-Gaussian CMB and LSS statistics beyond polyspectra}

\vskip 2.0cm

{\Large 
Gonzalo A. Palma$^\text{\scriptsize1}$, Bruno Scheihing H.$^\text{\scriptsize1}$ and Spyros Sypsas$^\text{\scriptsize1,2}$
}

\vskip 0.5cm

$^\text{\scriptsize1}${\it Grupo de Cosmolog\'ia y Astrof\'isica Te\'orica, Departamento de F\'{i}sica, FCFM, \mbox{Universidad de Chile}, Blanco Encalada 2008, Santiago, Chile  } \\ $^\text{\scriptsize2}${\it Department of Physics, Faculty of Science, Chulalongkorn University, Phayathai Rd., Bangkok 10330, Thailand}

\vskip 2.5cm

\end{center}

\begin{abstract} 

Cosmic inflation may have led to non-Gaussian initial conditions that cannot be fully parametrised by 3- and/or 4-point functions. In this work, we discuss various strategies to search for primordial non-Gaussianity beyond polyspectra with the help of cosmological data. Our starting point is a generalised local ansatz for the primordial curvature perturbation $\zeta$ of the form $\zeta = \zeta_{\rm G} + \F_{\rm NG} (\zeta_{\rm G})$, where $\zeta_{\rm G}$ is a Gaussian random field and $\F_{\rm NG}$ is an arbitrary function parametrising non-Gaussianity that, in principle, could be reconstructed from data. Noteworthily, in the case of multi-field inflation, the function $\F_{\rm NG}$ can be shown to be determined by the shape of tomographic sections of the landscape potential responsible for driving inflation. We discuss how this generalised local ansatz leads to a probability distribution functional that may be used to extract information about inflation from current and future observations. In particular, we derive various classes of probability distribution functions suitable for the statistical analysis of the cosmic microwave background and large-scale structure.

\end{abstract}

\end{titlepage}

\hypersetup{linktocpage}
%\hrulefill
\noindent\makebox[\linewidth]{\rule{\textwidth}{1.3pt}}
%\vspace{-20pt}
\tableofcontents
\noindent\makebox[\linewidth]{\rule{\textwidth}{1.3pt}}
%\hrulefill

\newpage

\setcounter{equation}{0}
\section{Introduction}

Our observable universe is consistent with an extremely simple set of initial conditions. For all practical purposes, the observed cosmological inhomogeneities were seeded by a primordial curvature fluctuation $\zeta$ distributed according to a Gaussian profile parametrised by an almost scale invariant power spectrum~\cite{Komatsu:2003fd, Aghanim:2018eyx, Akrami:2018odb, Akrami:2019izv, Akrami:2019bkn}. The confirmation of this state of affairs by future surveys would reinforce our confidence in the single-field slow-roll inflation paradigm, that is, the idea that $\zeta$ was the consequence of quantum perturbations of a single scalar fluid (the inflaton) that evolved adiabatically during inflation~\cite{Guth:1980zm, Starobinsky:1980te, Linde:1981mu, Albrecht:1982wi, Mukhanov:1981xt}. Tiny deviations from Gaussianity, due to small nonlinear self-interactions affecting $\zeta$, are known to emerge in single-field inflation but these are predicted to be too small to be observed in the near future. On the other hand, large non-Gaussianity (within current bounds) may emerge in models of inflation beyond the canonical single-field paradigm, resulting from possible nongravitational self-interactions of $\zeta$ and/or interactions between $\zeta$ and other degrees of freedom. The observation of non-Gaussianity (NG) would therefore offer a unique opportunity to characterise the class of interactions that affected $\zeta$ during inflation, allowing us to pin down certain fundamental aspects about the period of inflation and, consequently, have a glimpse on the structure of the ultra-violet (UV) framework where it is realised.

While current cosmic microwave background (CMB) observations show no evidence of primordial non-Gaussianity, future large-scale structure (LSS) surveys, such as {\sc Lsst}~\cite{Abell:2009aa}, {\sc Euclid}~\cite{Laureijs:2011gra}, {\sc Sphere}x~\cite{Dore:2014cca} and {\sc Ska}~\cite{Bull:2018lat}, promise to revitalise its search. The proliferation of modes due to the three-dimensional volume probed by galaxy surveys is expected to yield constraints on primordial non-Gaussianity that might not only complement current bounds from CMB observations, but even surpass them. Among the most prominent effects of non-Gaussianity on the matter distribution is the celebrated observation that a nonzero skewness of the probability distribution function of $\zeta$ leads to an enhanced abundance of collapsed structures and a scale dependent correction in the halo bias~\cite{Dalal:2007cu}, a result which has brought LSS surveys in the front line of the search for non-Gaussianity. Furthermore, there has been an intense effort to study how UV physics can show up in the matter power spectrum and bispectrum~\cite{Bartolo:2017sbu,MoradinezhadDizgah:2018ssw,Cabass:2018roz,MoradinezhadDizgah:2018pfo}. 

Given that inflation might not be unique\footnote{Other scenarios, such as the ekpyrotic~\cite{Lehners:2007ac,Khoury:2001wf} and bouncing universe~\cite{Wands:1998yp,Finelli:2001sr}, have been proposed as models able to reproduce nearly Gaussian initial conditions.} in explaining the observed inhomogeneities~\cite{Chen:2018cgg}, in order to confront future data with theory, we may adopt an agnostic perspective about the details involved in the description of the pre-Big-Bang dynamics. In that case, we should agree that the main outcome from inflation, or any other framework intending to explain the initial conditions of our universe, consists of a relation giving us back the profile of $\zeta (\x)$ written in terms of a purely Gaussian random field $\zeta_{\rm G}(\x)$. Such a relation must be of the form
\be \label{general-ansatz}
\zeta(\x) = \zeta_{\rm G}(\x) + \F_{\rm NG}\left( \zeta_{\rm G}(\x), \nabla \right),
\ee
where $\F_{\rm NG}$ represents a nonlinear function of the field $\zeta_{\rm G}(\x)$ and spatial gradients $\nabla$ acting on it. In Fourier space, the previous relation may be reexpressed as the following expansion in powers of $\zeta_{\rm G}$, starting at quadratic order:
\be \label{general-NG-k}
\zeta_\K = \zeta_\K^G + (2 \pi)^3 \sum_{n=2} \frac{1}{n!}\int_{\K_1} ... \int_{\K_n} \delta^{(3)} \Big(\K - \sum_{i=1}^n \K_i \Big) F_n\left( \K_1 , ... , \K_n \right)\zeta^G_{\K_1} ... \zeta^G_{\K_n}, 
\ee
where $\int_\K \equiv (2 \pi)^{-3} \int \de^3 k$, and where $F_n\left( \K_1 , ... , \K_n \right)$ are functions of the momenta, symmetric under their permutations. This relation is sufficiently general to describe any form of primordial non-Gaussianity, and may be formally obtained in a generic manner from a quantum field theoretical framework (see App.~\ref{sec:gla} for details). The $F_n$ functions parametrise the non-Gaussian deviations generated by nonlinear interactions to which $\zeta$ were subject and, in the case of inflation, may be deduced from a particular model by studying the evolution of $\zeta_\K$ from sub-horizon scales up until the end of inflation (\emph{e.g.}, using the in-in formalism). In fact, any $n$-point correlation function for $\zeta$ may be computed out of~(\ref{general-NG-k}). For instance, at tree-level, the bispectrum parametrising the amplitude of the 3-point function is found to be given by $B(\K_1 , \K_2 , \K_3) = \left[ P_{\zeta}(k_1) P_{\zeta}(k_2)   F_2 (\K_1, \K_2)  + \textrm{perm} \right]$, where $P_{\zeta}$ is the power spectrum of the Gaussian field $\zeta^G_{\K}$. In the particular case of single-field slow-roll inflation, up to first order in the slow-roll parameters, the bispectrum is recovered as long as $F_2$ is given by (up to terms that vanish upon imposing momentum conservation)
\be \nn
F_2 (\K_1, \K_2) = \frac{1}{2} (\eta - \epsilon) + \frac{\epsilon}{2} \frac{k_1 + k_2}{|\K_1 + \K_2|} + 2 \epsilon \frac{k_1^2 + k_2^2}{|\K_1 + \K_2| (k_1 + k_2 + |\K_1 + \K_2|)}, 
\ee
where $\epsilon$ and $\eta$ are the usual slow-roll parameters describing the steady evolution of the FLRW background during inflation. The effective field theory (EFT) of inflation approach~\cite{Cheung:2007st} to study models beyond the canonical single-field paradigm will also yield a specific form of $F_2 (\K_1, \K_2)$, in which the sound speed of $\zeta$ plays an important role. Moreover, the well-known local ansatz~\cite{Gangui:1993tt, Komatsu:2001rj, Acquaviva:2002ud, Maldacena:2002vr, Bartolo:2004if, Liguori:2010hx, Chen:2010xka, Wang:2013eqj}, 
related to the presence of multi-field dynamics, corresponds to another particular instance of this relation, where $F_2 = \frac{6}{5} f_{\rm NL}^{\rm local}$.
 
One could thus take upon the challenge of directly reconstructing the form of $\F_{\rm NG}$ in Eq.~(\ref{general-ansatz}) ---or equivalently, the functions $F_n$ appearing in (\ref{general-NG-k})--- out of cosmological data. This would constitute a bottom-up approach to determine the properties of the model that gave origin to the initial conditions. To guide such a reconstruction, one could consider restricting the functions $F_n$ according to certain rules dictated by the symmetries of the alleged bulk model that led to~(\ref{general-NG-k}) at the end of the pre-Big Bang period. For instance, scale invariance of the spectra is equivalent to the invariance of $F_n$ under the simultaneous rescaling of all the momenta: $F_n (\lambda \K_1 , ... , \lambda \K_n) = F_n ( \K_1 , ... ,  \K_n)$. Furthermore, the validity of soft theorems~\cite{Maldacena:2002vr, Creminelli:2004yq, Cheung:2007sv, Tanaka:2011aj, Creminelli:2012ed, Assassi:2012zq, Pajer:2013ana, Joyce:2014aqa, Bravo:2017wyw, Finelli:2017fml, Bravo:2017gct} (under certain circumstances) would require some relations among $F_n$ of different order in the limit where one or more of the momenta vanish.

An objection to this program (the direct reconstruction of $\F_{\rm NG}$) is that perturbation theory applied to the study of the evolution of $\zeta$ implies that the expansion involved in the writing of Eq.~(\ref{general-NG-k}) is hierarchical. That is, given a small dimensionless coupling constant $g$ parametrising the self-interactions experienced by $\zeta$ during inflation, the $F_n$ functions are naturally expected to satisfy
\be \label{fn-intro}
F_n \propto g^{n-1}.
\ee 
For instance, in the case of single-field slow-roll inflation $g$ happens to be of order $\epsilon$ and $\eta$ and hence, non-Gaussianity is expected to be well-parametrised by the bispectrum. In noncanonical single-field models described by the EFT of inflation, where the $\zeta$ fluctuations propagate with a reduced sound speed $c_s < 1$, the coupling $g$ is enhanced by a factor $c_s^{-2}$, but in order to trust perturbation theory, one still requires that it stay sufficiently suppressed. Based on this argument, we could say that most efforts to characterise non-Gaussianity so far have focused on a truncated version of~(\ref{general-NG-k}), where only $F_2 ( \K_1 , \K_2 )$ and $F_3 ( \K_1 , \K_2 , \K_3 )$ (which at tree-level give the bispectrum and trisprectrum) are taken under consideration, with the {\sc Planck} data implying weak constraints on the form of $F_2 ( \K_1 , \K_2 )$, with the help of the so-called local, equilateral, orthogonal and folded templates~\cite{Akrami:2019izv}.

In the present article, we wish to argue in favour of reconstructing the full function $\F_{\rm NG}$ of Eq.~(\ref{general-ansatz}) from CMB and LSS observations without necessarily assuming a hierarchical dependence of the functions $F_n$ on a given coupling constant $g$. We posit that the search for non-Gaussianity focused on low $n$-point correlation functions may miss the existence of richer types of statistics~\cite{Bond:2009xx,Leblond:2010yq, Suyama:2013dqa,Flauger:2016idt, Chen:2018uul, Chen:2018brw,Panagopoulos:2019ail} that we may be unable to predict by following standard perturbation theory techniques. For example, in App.~\ref{sec:gla} we show that in certain classes of multi-field models, the function $\F_{\rm NG}$ is found to be proportional to the gradient of the landscape potential $\Delta V(\psi)$ controlling the dynamics of the isocurvature field $\psi$. More to the point, we show that 
\be
\F_{\rm NG} (\zeta) \propto \frac{\partial }{\partial \zeta} \Delta V \left(\psi (\zeta) \right)  , \qquad  {\rm with}  \qquad   \psi (\zeta)\equiv \frac{H}{2\pi A_s^{1/2}} \zeta ,
\ee
where $H$ is the Hubble expansion rate during inflation and $A_s$ is the amplitude of the power spectrum of $\zeta$ (constrained by {\sc Planck} to be $10^{9} A_s = 2.105 \pm 0.030$ at the CMB pivot scale $k = 0.05$ Mpc$^{-1}$~\cite{Aghanim:2018eyx}). Hence, the shape of $\F_{\rm NG}$ provides a tomographic view of the landscape potential away from the inflationary trajectory.
Certainly, one cannot discard the possibility of having potentials $\Delta V (\psi)$ characterised by a rich structure (\emph{i.e.} with consecutive minima separated by field distances smaller than the Hubble expansion rate during inflation) in which case the bispectrum would not constitute an efficient tool to parametrise non-Gaussianity~\cite{Chen:2018uul,Chen:2018brw}. A simple example of this, related to the presence of an axionic isocurvature direction, is $\F_{\rm NG} (\zeta_{\rm G})  \propto \sin (\zeta_{\rm G} / f_\zeta)$ with $f_\zeta^2 < \sigma_\zeta^2$, where $\sigma_\zeta^2$ is the variance of the Gaussian field $\zeta_{\rm G}$. 
In this case, the function $\F_{\rm NG} (\zeta_{\rm G})$ remains bounded, but varies vigorously within the relevant range determined by $\sigma_\zeta$.

Despite of having a well motivated construction justifying the need of studying non-Gaussianity beyond polyspectra, we wish to keep the discussion as general as possible and, thus, in the main part of this work we will not refer to any specific origin of this parametrisation. Instead, we will focus our attention on the consequences ---for various cosmological observables--- implied by having initial conditions parametrised by a function $\F_{\rm NG} (\zeta)$ not necessarily respecting a hierarchical structure. In order to make the discussion more tractable, we will focus on the specific case of local NG, for which $\F_{\rm NG}$ in Eq.~(\ref{general-ansatz}) does not involve gradients and hence the coefficients $F_n$ in the expansion~(\ref{general-NG-k}) are independent of momenta\footnote{See Ref.~\cite{dePutter:2016moa} for a discussion on how the usual local ansatz involving $f_{\rm NL}$ (where the momenta are disregarded) can in principle be observationally distinguished from an ansatz where the function $F_2$ satisfies the consistency relation. The methods used in~\cite{dePutter:2016moa} should be valid to analyse our proposed generalised version of the local ansatz.}. This implies a simplified version of~(\ref{general-ansatz}) given by $\zeta(\x) = \zeta_{\rm G}(\x) + \F_{\rm NG}\left( \zeta_{\rm G}(\x) \right)$. However, since in practice observations are restricted to a finite range of wavelengths, we will consider a version of~(\ref{general-ansatz}) recovered from (\ref{general-NG-k}) but with every momenta restricted to a finite range (including those appearing in the integrals). This relation is given by
\be \label{general-ansatz-restricted}
\zeta(\x) = \zeta_{\rm G}(\x) + \F_{\rm NG}\left[ \zeta_{\rm G} \right](\x) , \qquad {\rm with } \qquad \F_{\rm NG}\left[ \zeta_{\rm G} \right](\x) =  \int_{\K} \int_\y e^{i\K \cdot (\x - \y) } F\left( \zeta_{\rm G}(\y) \right),
\ee
where $\int_\y \equiv  \int \de^3 y$. In the previous expression, $\int_{\K}\equiv (2\pi)^{-3}  \int \de^3 k$ is restricted to a given range of scales specified by convenience\footnote{For now, we do not need to specify the range of momenta and the reader may take the range as the entire momentum space, in which case $\F_{\rm NG} = F$. In Sec.~\ref{sec:PDF-run-ren}, we will consider the problem of reducing the range of momenta (in a Wilsonian manner) and explore how the function $F$ appearing in Eq.~(\ref{general-ansatz-restricted}) runs with the IR and UV cutoffs employed to restrict the momenta.} (with $\zeta(\x)$ and $\zeta_{\rm G}(\x)$ also restricted in that manner).  In this version of the ansatz, $F\left( \zeta_{\rm G} \right)$ is the function of the Gaussian random field $\zeta_{\rm G}$ that determines the values of the $F_n$ coefficients in (\ref{general-NG-k}) as $F_n = \partial_{\zeta}^n F |_{\zeta = 0}$. 

While similar approaches to study non-Gaussianity have been developed (for instance, see Refs.~\cite{Bond:2009xx,Suyama:2013dqa}, where the main focus is the analysis of non-Gaussianities arising from preheating), our purpose in this work is to provide a description of local non-Gaussianity in terms of the curvature perturbation field $\zeta$ in a way that explicitly reflects how the scales involved in the computation affect the result, and therefore, on how the primordial deviation from Gaussianity $\F_{\rm NG}$ can be related to predictions and observables in the CMB and LSS by adjusting the set of scales to those involved in the measurement at hand.

\subsection{Main results} 

\renewcommand{\labelitemi}{$\star$}

The object that determines the statistics of fluctuations, either in the temperature or in the density field, is the probability distribution function (PDF) from which one can compute moments, cumulative functions and several other objects such as mass functions, bias factors, etc. Since both the temperature and density fluctuations are sourced by curvature perturbations, the parametrisation (\ref{general-ansatz-restricted}) of $\zeta$ in terms of Gaussian random fields allows us to construct various classes of such functions characterising the statistics\footnote{There are several nonprimordial sources of non-Gaussianity, especially in late-time fields. Here, we always refer to its primordial component; see Sec.~\ref{sec:scnd-order-pt} for comments on the limitations posed by secondary effects.} of CMB and LSS, hence, enabling us to use their distributions as a probe of the primordial universe. In this work we derive several classes of PDFs for the curvature perturbation $\zeta$, the CMB temperature fluctuation $\Theta \equiv \Delta T/T$ and the matter overdensity $\delta \equiv \delta \rho / \rho$, all following from the non-Gaussian ansatz \eqref{general-ansatz-restricted} by using appropriate transfer functions. 

Specifically, we achieve the following goals:
\begin{itemize}
\item We derive the full probability distribution functional implied by the generalised ansatz (\ref{general-ansatz-restricted}). The result is shown in Eq.~\eqref{PDF-windowed} and it explicitly depends on the function $F$ (which at the same time, is related to the landscape potential in the case where the primordial non-Gaussianity is produced during inflation). 
\item We use this functional to derive various statistical tests for CMB and LSS, among them, 1- and 2-point PDFs, written in Eqs.~\eqref{one-point} and~\eqref{two-point}. We also write the corresponding PDFs for the CMB temperature anisotropies in Eqs.~\eqref{one-point-T} and \eqref{two-point-th}.
\item We derive the NG halo mass function expressed in terms of the function $F$. This result is shown in Eqs.~\eqref{mass-function-PS-F}-\eqref{mass-function}. We further comment on how statistical estimators probing the PDF can be used in LSS datasets in connection to primordial NG via Eq.~\eqref{mu-dn}.
\item We derive a generalised version of the linear halo bias resulting from the non-Gaussian deformation~$\F_{\rm NG}$. The result is written in Eq.~(\ref{linear-b-1}), where we show that the scale dependent bias is sensitive to the full function $F$ and not to any specific truncated version of it. In other words, all nonlinearity parameters appear in a degenerate manner.
\end{itemize}
To summarise, we wish to point out that since non-Gaussianity has not shown up in low $n$-point functions and given that its presence has solid foundations in our understanding of UV physics and quantum field theory/gravity, extending our study towards other estimators is part of the next step. The PDF is the object that fully encodes the statistical information of fluctuations enclosing all possible cases and as such it may serve as a central tool in the search for non-Gaussianity.

\subsection{Structure of the paper} 

The article is organised as follows: In Sec.~\ref{sec:zeta-stat}, we write down the probability functional describing the distribution of the curvature fluctuation $\zeta$, its Fourier dual ---the partition function--- as well as the 1-point and 2-point PDFs. We discuss several aspects regarding the use of window functions related to renormalisation, survey sky coverage and secondary NG effects. In Sec.~\ref{sec:theta-stat}, we derive the temperature distribution and focus on the 2-point PDF commenting on its use with CMB data. In Sec.~\ref{sec:delta-stat}, we derive the halo mass function and the halo bias for arbitrary local NG. We conclude in Sec.~\ref{sec:concl}, while in the Appendices we present various details regarding the NG ansatz and the functionals emanating thereof presented in the main text.

\setcounter{equation}{0}

\section{$\zeta$ Statistics} \label{sec:zeta-stat}

 In this section we derive the probability distribution functional (and other classes of distributions) for the primordial curvature perturbation $\zeta$. Our starting point is the generalised local ansatz shown in Eq.~\eqref{general-ansatz-restricted}. The resulting PDF's will set the stage for later sections, where we explore the consequences their non-Gaussian deviations may have on both CMB and LSS datasets.

\subsection{Probability distribution functional: ``bare'' theory}

The first task we can perform, and the most readily available to us thus far, is to attempt to use Eq.~\eqref{general-ansatz-restricted} to its full potential and derive directly the complete functional distribution that governs the $\zeta$ statistics: given that we know how a Gaussian random field is distributed, we may simply perform a change of variables to obtain the probability distribution functional $\PDF [\zeta]$ for $\zeta(\x)$ such that
\be
\langle \zeta(\x_1) \cdots \zeta(\x_n)\rangle = \int \! D \zeta \, \PDF [\zeta] \, \zeta(\x_1) \cdots \zeta(\x_n) ,
\ee
where $D\zeta$ is the functional measure of the PDF. Such an approach was used in~\cite{ColesBarrow} for the study of certain non-Gaussian initial conditions from inflation, in~\cite{Coles} in the context of nonlinear evolution of the correlation function of matter overdensities, while in~\cite{Matarrese:2000iz,Verde:2013gv} it was shown how initial conditions with nonzero skewness/kurtosis can be described by the change of variables from a Gaussian PDF. Here, we extend this technique to the case of the generic NG field of Eq.~\eqref{general-ansatz-restricted}.

The resulting distribution must account, to leading order in the perturbation $\F_{\rm NG}$, for every conceivable correlation function that may be constructed from the field $\zeta(\x)$ and for every expectation value of function(al)s of $\zeta(\x)$. To start with, the Gaussian random field $\zeta_{\rm G}$ is drawn from the following functional distribution:
\be
\PDF_{\rm G}[\zeta_{\rm G}] = \mathcal{N} \exp \left[-\frac{1}{2} \int_\x \int_\y \zeta_{\rm G}(\x) \Sigma^{-1}(\x,\y) \zeta_{\rm G}(\y) \right]= \mathcal{N} \exp \left[-\frac{1}{2} \int_\K \frac{\zeta_{\rm G}(\K) \zeta_{\rm G}(-\K)}{P_\zeta(k) }  \right] ,
\ee
where $\mathcal{N}$ is an overall normalisation constant, while $\Sigma^{-1}(\x,\y)$ and $P_{\zeta}(k)$ are the covariance matrix and the power spectrum respectively, related as
\be \label{power-cov}
\Sigma^{-1}(\x,\y) \equiv \int_\K \frac{e^{i \K \cdot (\x-\y) } }{P_{\zeta}(k)}.
\ee
In implementing the transformation $\zeta(\x) = \zeta_{\rm G}(\x) + \F_{\rm NG}[\zeta_{\rm G}](\x)$, there are two contributions that emerge: one is composed by the terms that come out of the Taylor expansion of the Gaussian distribution by regarding the perturbation $\F_{\rm NG}$ as small, and the other arises from the functional determinant of the transformation.  The latter is given by
\bea
{\rm det}\left( \frac{\delta \zeta(\y) }{\delta \zeta_{\rm G}(\x) } \right) = \exp \left[ {\rm tr} \left( \ln \left( \frac{\delta \zeta(\y) }{\delta \zeta_{\rm G}(\x) } \right) \right) \right] &=&   \exp \left[ {\rm tr} \left( \int_{\K} e^{i \K \cdot (\x - \y)} \ln \left\{ 1 + \frac{\de F}{\de \zeta}(\zeta_{\rm G}(\x)) \right\} \right) \right] \nonumber \\
&=& \exp \left[ \int_\x \int_{\K} \ln \left\{ 1 + \frac{\de F}{\de \zeta}(\zeta_{\rm G}(\x)) \right\} \right].
\eea
As long as the determinant exists (concretely, $\de F/\de\zeta > -1$) and is nonzero, we may, in principle, find an inverse to the relation $\zeta = \zeta_{\rm G} + \F_{\rm NG}[\zeta_{\rm G}]$ and denote it with $\zeta_{\rm G}(\x) = G[\zeta](\x)$. Then we may change variables from the PDF associated to $\zeta_{\rm G}$, to find
\be \label{exact-PDF}
\PDF[\zeta] = \mathcal{N} \exp \left[ -\frac{1}{2} \int_\x \int_\y G[\zeta](\x) \Sigma^{-1}(\x,\y) G[\zeta](\y) -  \int_\x \int_{\K} \ln \left\{ 1 + \frac{\de F}{\de \zeta}(\zeta_{\rm G}(\x)) \right\} \right].
\ee
This is an exact result. However, if $F$ is small in comparison to the typical scales of the background theory on which the fluctuation field $\zeta$ lies, and so is $\de F/\de\zeta$, we can approximate the logarithm in the exponential with the first term in its power series expansion, and furthermore, we may approximate the inverse mapping by $G[\zeta] \approx \zeta - \F_{\rm NG}[\zeta]$. This yields a non-Gaussian exponential correction to the PDF, the exponent of which, to first order in $\F_{\rm NG}$, reads
\be \label{exp-PDF-1}
\PDF[\zeta] = \PDF_{\rm G}[\zeta] \times \exp \left[ \int_\x \int_\y \zeta(\x) \Sigma^{-1}(\x,\y) \F_{\rm NG}(\zeta(\y))  -  \int_\x \frac{\delta}{\delta \zeta(\x)} \int_{\K} \int_\y e^{i \K \cdot (\x - \y)}  \F_{\rm NG}(\zeta(\y)) \right].
\ee
Using the definition of the covariance~\eqref{power-cov}, we may finally write the result as
\be \label{exp-PDF-2}
\PDF[\zeta] = \PDF_{\rm G}[\zeta] \times \exp \left[ \int_\x \int_{\K} \int_\y e^{i \K \cdot (\x - \y) } \left(  \frac{\zeta(\x)}{P_\zeta(k)} - \frac{\delta}{\delta \zeta(\x)}  \right) \F_{\rm NG}[\zeta](\y) \right].
\ee
This functional will serve as the guiding principle for all subsequent results. Note that even in a first order approximation, the probability distribution is always positive. However, to make computations tractable, we find it convenient to also perform a power series expansion of the non-Gaussian exponential factor. Then, to first order in $\F_{\rm NG}$, we find that 
\bea \label{PDF}
\PDF_F [\zeta] &=& \PDF_{\rm G}[\zeta] \times  \left[ 1 -  \int_\x \int_\y \int_\K e^{ i \K \cdot (\x - \y)}  \left( \frac{\delta}{\delta \zeta(\x)}  - \frac{\zeta(\x)}{P_\zeta(k)}  \right) \F_{\rm NG}[\zeta](\y) \right] \nonumber \\
&=& \PDF_{\rm G}[\zeta] \times \left[ 1 -  \int_\y \int_\K e^{-i \K \cdot \y}  \left( \frac{\delta}{\delta \zeta(-\K)}  - \frac{\zeta(\K)}{P_\zeta(k)}  \right) \F_{\rm NG}[\zeta](\y) \right].
\eea
The functional distribution~\eqref{PDF} has support at all the scales where the underlying theory does, or at least, at the scales where the corresponding EFT is presumed to hold true. However, observable quantities do not typically involve all of the scales, and therefore it may be that the ``bare'' departure from Gaussianity $\F_{\rm NG}$ is not the most adequate quantity to describe them. We thus now turn to the discussion of using window functions and how to integrate scales out.

\subsection{Probability distribution functional: running and renormalising} \label{sec:PDF-run-ren}

When making predictions, any EFT will force us to recognise certain scales at which the theory is no longer well-suited to describe physical observables. This typically implies a high-energy scale, where the theory has to be cut off. Thus, we will set $k_{\rm UV}$ as the maximum possible wavenumber the mode expansion of the curvature perturbation can have. Similarly, while it is not always introduced, one can make the same assertion with the very long wavelengths. As much as our theory may have predictions concerning phenomena happening beyond the present Hubble radius, they are currently unobservable. Therefore, in establishing predictions for observable quantities it seems natural to integrate out those scales, so that they are properly incorporated into the final, effective result. Because of this, we will take a conservative attitude and also define an infrared cutoff $k_{\rm IR}$, which can be thought of as the inverse of the current Hubble radius, thus bounding the domain of the theories we will be studying to $k \in (k_{\rm IR}, k_{\rm UV})$.

Moreover, in realistic situations the experiment at hand may not allow us to access every value for the momentum scale $k$ evenly. In those cases we may wish to introduce a window function $W(k)$ to filter our results and give more weight to some scales. Accordingly, one would be interested in the statistics of the filtered field
\be \label{z-filtered}
\zeta_W (\x) \equiv  [W \star \zeta](\x) = \int_\K \int_\y e^{i{\K} \cdot (\x - \y)} W(k) \zeta(\y).
\ee
To derive the probability distribution functional of the field $\zeta_W (\x)$ it is enough to perform the change of variables $\zeta \to \zeta_W({\K} ) = W(k) \zeta({\K})$ in~(\ref{PDF}). One obtains
\be \label{PDF-windowed}
\begin{split}
\PDF_{W}[\zeta_W] = & \mathcal{N}_W  \exp \left[ -\frac{1}{2} \int_\K \frac{\zeta_W(\K) \zeta_W(-\K)}{P_W(k) } \right] \\ & \times  \left[ 1 -  \int_\y \int_\K  e^{-i \K \cdot \y} W(k) \left( \frac{\delta}{\delta \zeta_W(-\K)}  - \frac{\zeta_W(\K)}{P_W(k)}  \right) \F_{\rm NG} \left[ \int_q  \frac{e^{-i\q \cdot \y} \zeta_W(\q)}{W(q)} \right](\y) \right],
\end{split}
\ee
where $P_W(k) \equiv W^2(k) P_\zeta(k)$. Leaving aside the argument of the deviation from Gaussianity $\F_{\rm NG}$ for a moment, this PDF has the same structure as the unfiltered PDF of Eq.~(\ref{PDF}). 

This expression for $\PDF_{W}$ poses an interesting question: what if the window function of choice is defined (as usual) with hard cutoffs, just as if we were redefining the limits of our EFT? That is to say, how does $\PDF_{W}$ look if we have
\be \label{flat}
W(k) = \begin{cases}
1 & \text{if} \,\, k \in (k_L, k_S) \\
0 & \text{if} \,\, k \not\in (k_L, k_S),
\end{cases}
\ee
as the window function? It turns out that, for the functional integral to be well-defined, we have to integrate out of the theory all the modes which will not take part in our observable quantities. Upon doing so, we can avoid dealing with $\zeta_W(\q)/W(\q)$, which from the perspective of the theory with the window function would be ill-defined for the scales where $W=0$, but is perfectly finite (and equal to $\zeta(\q)$) from the perspective of the original theory. To this end, let us take the original functional distribution~\eqref{PDF} (with $k$-space variables) and integrate out the modes outside the support of the window function $W$. We may write this as 
\be \label{ren-PDF}
\PDF_{W}[\zeta] = \int D\zeta_{|\K| \notin (k_L, k_S)} \PDF[\zeta].
\ee
Upon integration over the prescribed range of modes, the purely Gaussian term in~\eqref{PDF} gives a reduced Gaussian measure that considers only $k \in (k_L, k_S)$. To integrate the term containing $\F_{\rm NG}$ it is convenient to separate the integral over momentum space in~\eqref{PDF} into two contributions:
\be \label{in-out}
\begin{split}
\int_\K e^{-i \K \cdot \y}  \left( \frac{\delta}{\delta \zeta(-\K)} - \frac{\zeta(\K)}{P_\zeta(k)}  \right) \F_{\rm NG}[\zeta](\y) &= \int_{|\K| \in (k_L,k_S)} e^{-i \K \cdot \y}  \left( \frac{\delta}{\delta \zeta(-\K)}  - \frac{\zeta(\K)}{P_\zeta(k)}  \right) \F_{\rm NG}[\zeta](\y) \\ & + \int_{|\K| \not\in (k_L,k_S)} e^{-i \K \cdot \y}  \left( \frac{\delta}{\delta \zeta(-\K)}  - \frac{\zeta(\K)}{P_\zeta(k)}  \right) \F_{\rm NG}[\zeta](\y).
\end{split}
\ee
In this form, it is easy to see that upon integrating over $\zeta(\K)$, the term in the second line of (\ref{in-out}) vanishes after performing a functional partial integration. Thus, we will only need to know how to deal with the quantity\footnote{Note that when considering the functional integration of the first line of Eq.~\eqref{in-out}, the differential operator $\left( \frac{\delta}{\delta \zeta(-\K)}  - \frac{\zeta(\K)}{P_\zeta(k)}  \right)$ involves scales $|\K| \in (k_L, k_S)$, while the functional integration goes over modes with $|\K| \notin (k_L, k_S)$. We may thus pull it out of the integral.}
\be \label{renorm}
 \left( \frac{\delta}{\delta \zeta(-\K)}  - \frac{\zeta(\K)}{P_\zeta(k)}  \right) \int D\zeta_{|\K| \notin (k_L, k_S)} \exp \left[-\frac{1}{2} \int_\K \frac{\zeta(\K) \zeta(-\K)}{P_\zeta(k)}  \right] \int_\y \int_{|\K| \in (k_L, k_S)} \!\!\!\!\!\!\!\! e^{-i \K \cdot \y} \F_{\rm NG}[\zeta](\y),
\ee
so as to see if, and how, the interaction is renormalised. It turns out that if we define 
\be \label{def-bar-F}
{\bar F}(\zeta_W(\x)) \!\equiv\! \int D\zeta_{k \notin (k_L, k_S)} \exp \left[-\frac{1}{2} \int_{|\K| \notin (k_L,k_S)} \!\!\!\!\!\! \frac{\zeta(\K) \zeta(-\K)}{P_\zeta(k)}  \right] F(\zeta(\x)) = \int_{-\infty}^{\infty} \!\!\de{\bar \zeta } \frac{e^{-\frac{(\zeta_W(\x) - {\bar \zeta})^2 }{2 \sigma_{\rm out}^2 }  }  }{\sqrt{2 \pi \sigma_{\rm out}^2 } } F(\bar \zeta),
\ee
then we may identify 
\be
\bar \F_{\rm NG}[\zeta](\x) = \int_{\y} \int_{|\K| \in (k_L, k_S)} \!\!\!\!\!\!\!\! e^{i \K \cdot (\x - \y) } \bar F(\zeta(\y) ) ,
\ee
as the effective self-interaction, because we would have integrated out all the scales that are outside the range of interest and still leave the other scales within the measure $\PDF$ of the PDF, while maintaining its analytic structure.

The last equality of~\eqref{def-bar-F}, the Weierstrass transform of $F(\zeta)$, can be obtained in numerous manners. If one were to follow standard diagrammatic perturbation theory, it arises from summing back every ``loop'' contraction performed by the Gaussian measure of $\F_{\rm NG}$ with itself. Since the momenta flowing through those loops are bounded by the range being integrated out, and there is no ``external'' momenta flowing through the diagrams, we have that their numerical value is the same for every loop and equal to the variance
\be
\sigma_{\rm out}^2 = \int_{ |\K| \not\in (k_L, k_S)} \!\!\!\!\!\! P_\zeta(k),
\ee
that was subtracted from the Gaussian field statistics when the modes $|\K| \not\in (k_L, k_S)$ were integrated out of the theory. Therefore, by using these results in Eq.~\eqref{ren-PDF}, we obtain
\be
\PDF_{\F_{\rm NG},W}[\zeta] = \PDF_{\bar \F_{\rm NG}}[\zeta_{|\K| \in(k_L, k_S)}],
\ee
\emph{i.e.}, that $\PDF_{W}[\zeta]$ for the restricted variable~\eqref{z-filtered} has the same functional form as the original PDF, with the only modification that now the departure from Gaussianity is given by a ``filtered'' interaction $\bar \F_{\rm NG}$ instead of the ``bare'' interaction $\F_{\rm NG}$.

In practice, there is more than one way of how to represent the running of $F$ depending on the scales one wants to include in the theory. Perhaps the most ethereal representation, but at the same time the most revealing of the theory's structure, is through the differential expression of the Weierstrass transform, as in
\be \label{W-ren-out}
{\bar F}(\zeta) = \exp \left[ \frac{\sigma_{\rm out}^2 }{2} \frac{\partial^2}{\partial \zeta^2 } \right] F(\zeta).
\ee
This expression makes clear how the theory runs by removing more or less scales, as well as the fact that the transformation rule between $F$ at different scales follows an adequate composition property: integrating out ranges of momenta $A$ and $B$ is implemented via $\sigma_{\rm out,A}^2$ and $\sigma_{\rm out,B}^2$, and doing so yields the same result independently of the order in which one subtracts the modes from the theory. Furthermore, this shows that the functional form of $\F_{\rm NG}$ and $\bar \F_{\rm NG}$ is the same, in the sense that the quantities that determine its concrete expression are exactly the same; the only thing that the window does is to restrict the range of modes entering in the observables. 

Conversely, just as the PDF may be recast in an analogous manner to that of the original theory, the field with modes between $k_L$ and $k_S$ may also be written down as a local departure from Gaussianity
\be \label{local-windowed}
\zeta_W(\x) = \zeta^G_W(\x) + {\bar \F_{\rm NG}}\left[ \zeta_W^G \right](\x),
\ee
merely because it follows statistics analogous to $\zeta$. Here we have to remind the reader that this is only so for $W(k)$ of the form~\eqref{flat}. Other window functions still give rise to an explicit PDF, namely equation~\eqref{PDF-windowed}, but the deviations from Gaussianity may no longer be written as concisely as in~\eqref{local-windowed}. The difference lies in that a general window function does not render irrelevant some degrees of freedom of the theory; it only gives them dissimilar weights in the final result. However, in order to obtain the function ${\bar \F_{\rm NG}}$ it is crucial that we reduce the number of independent variables in our theory by integrating out their effects, as all of them will leave their signature, if small, in any given correlation function. Having made this point explicit, from now on, and for the rest of this paper, we will go back to considering general window functions.

Thus far, we have established how we may write the probability distribution functional of our theory depending on the scales under consideration and also how we may incorporate window functions into the distribution functional. Before passing to simpler statistical estimators stemming from this functional, we now discuss the partition function.

\subsection{Partition function and $n$-point correlators}

In practice, using the full probability distribution functional directly on cosmological data proves to be difficult, as there is only one realisation of our universe to probe and conduct measurements in, so it is not possible to take a frequentist approach to its statistics. While this suggests the use of Bayesian analysis to find the most probable $\F_{\rm NG}$ given the data by the means of the functional~\eqref{PDF}, it also reveals why one typically chooses to work with correlation functions to probe departures from Gaussianity: they can be computed from many Fourier modes on the sky, whose past history is presumably independent (at least if the nonlinearities are turned off), and therefore averages may be performed and compared with the theoretical predictions for the expectation values or correlations.

Fittingly, there is an object that encapsulates the information of all the correlation functions in a perhaps clearer way than the full probability density functional $\PDF$. This is the partition function $Z[J]$, which is the object that generates the $n$-point functions via functional differentiation:
\be \label{Z-corr}
\langle \zeta({\K}_1 ) \cdots \zeta({\K}_n )\rangle = \frac{\delta^n Z[J]}{(i\delta J(-{\K}_1 )) \cdots (i \delta J(-{\K}_n)) }\bigg|_{J = 0},
\ee
or equivalently, the functional Fourier transform of the PDF, which in the context of probability theory is called the characteristic function,
\be \label{Z-Fourier}
Z[J] = \int D \zeta \, \PDF[\zeta] \, e^{i\int_\K \zeta({\K}) J(-{\K})}.
\ee

Both expressions may be employed to obtain $Z[J]$: the first requires to know all the $n$-point functions beforehand and reconstruct the object that has them as its functional derivatives, while the second requires to know an explicit expression for the probability distribution functional. Since we have the latter, we may carry out this computation explicitly\footnote{The details of this derivation are presented in App.~\ref{sec:Partition}.} to first order in $\F_{\rm NG}$, obtaining
\be \label{partition}
\begin{split}
Z[&J]  = \, \exp \left[-\frac{1}{2} \int_\K J(\K) J(-\K) P_\zeta(k) \right] \;\times\\  &  \left( 1 - \int_\x \frac{\int_\K e^{i\K \cdot \x} J(-\K)}{ \int_\K e^{i\K \cdot \x} J(-\K) P_\zeta(k) } \int_{\bar \zeta} \frac{\exp \left[ - \frac{\left({\bar \zeta} - i\int_\K e^{i\K \cdot \x} J(-\K) P_\zeta(k) \right)^2}{2 \sigma_\zeta^2} \right] }{\sqrt{2 \pi} \sigma_\zeta} \left(\sigma_\zeta^2 \frac{\partial }{\partial \bar \zeta} - \bar{\zeta}\right)  F(\bar \zeta) \right).
\end{split}
\ee
Here we have defined $\sigma_\zeta^2 \equiv \int_k P_\zeta(k)$ as the 1-point variance associated to the power spectrum for the relevant range of momenta. Window functions are easily incorporated by substituting $J(\K)$ with $J(\K) W(k)$, as this procedure will add a factor of $W(k)$ to every external leg in any given diagram.

Now that we have equation~\eqref{partition}, we may compute the $n$-point functions directly, without having to resort to functional integration as we would with~\eqref{PDF}. Moreover, the structure that will emerge in these correlations is more closely related to~\eqref{partition}, as is demonstrated by their explicit expressions in position space
\be \label{npt-position}
\langle \zeta_W({\x}_1 ) \cdots \zeta_W({\x}_n )\rangle_c = f_{n-1} \sum_{i=1}^n \int_{\x}  \frac{\int_{\K} W(k) e^{i\K \cdot (\x_i - \x) } }{\int_\K e^{i\K \cdot (\x_i - \x) } W(k) P_\zeta(k)  } \left( \prod_{i=1}^n \int_\K W(k) P_\zeta(k) e^{i \K \cdot (\x_i - \x) } \right),
\ee
where the subscript $c$ indicates the result only considers the fully connected piece. For completeness, we write down their counterparts in momentum space 
\be \label{n-pt-f}
\langle \zeta_W({\K}_1 ) \cdots \zeta_W({\K}_n )\rangle_c = f_{n-1} \, (2\pi)^3 \delta^{(3)} \left( \sum_{i=1}^n \K_i \right)  \left( \prod_{j=1}^n W(k_j) P_\zeta(k_j)  \right) \sum_{i=1}^n \frac{1}{P_\zeta(k_i)} , 
\ee
where the coefficients\footnote{In a more standard notation, the first few terms would correspond to $f_2=f_{\rm NL}$, $f_3=g_{\rm NL}$, etc.} $f_n$ are given by (probabilists') Hermite moments of $F$:
\be
f_n \equiv - \frac{1}{\sigma_\zeta^{n}} \int_{-\infty}^{\infty} \de\zeta \frac{e^{-\frac{\zeta^2}{2\sigma_\zeta^2 } } }{\sqrt{2\pi  }\sigma_\zeta } {\rm He}_{n}\! \left( \frac{\zeta}{\sigma_\zeta} \right) F (\zeta).
\ee
The coefficients $f_n$ are quantities of mass dimension $1-n$, which are invariant under the renormalisation procedure, discussed in Sec.~\ref{sec:PDF-run-ren}, in a very fitting sense: because $\{f_n\}_{n=2}^{\infty}$ are the coefficients of a Hermite polynomial expansion\footnote{We omit $n=0,1$ in the Hermite expansion because we assume $\langle \zeta \rangle = 0$ and $\langle \zeta \zeta \rangle $ to be set by the free theory matching the power spectrum of observations. To put it differently, with this definition of the local ansatz, due to orthogonality properties of the Hermite polynomials, the power spectrum of $\zeta$ is not modified to first order in the nonlinearity parameters~\cite{Smith:2011ub}.}, we have
\be \label{F-herm}
F(\zeta;\sigma_\zeta^2) = - \sum_{n=2}^{\infty} \frac{f_n}{n!} \, \sigma_\zeta^{n} \, {\rm He}_n \! \left( \frac{\zeta}{\sigma_\zeta} \right) = - \sum_{n=2}^{\infty} \frac{f_n}{n!} \exp \! \left[ - \frac{\sigma_\zeta^2}{2} \frac{\partial^2 }{ \partial \zeta^2} \right] \zeta^n,
\ee
where we have introduced the variance $\sigma_\zeta^2$ as an argument of $F$ in order to emphasise that the associated field has the corresponding amplitude for its fluctuations. This means that $\bar F$ in Eq.~\eqref{def-bar-F}, or equivalently Eq.~\eqref{W-ren-out}, takes the following form:
\be
{\bar F}(\zeta) = \exp \! \left[  \frac{\sigma_{\rm out}^2}{2} \frac{\partial^2 }{ \partial \zeta^2} \right] F(\zeta;\sigma_\zeta^2) = F(\zeta;\sigma_\zeta^2 - \sigma_{\rm out}^2),
\ee
where the coefficients $\{f_n\}_{n}$ remain unchanged; only the variance gets reduced to its new value after integrating some modes out.

Nonetheless, it is important to point out that each individual $n$-point function does not encapsulate all of the non-Gaussian information contained in $F$. Indeed, each correlation function only yields one term of an infinite series expansion of $F$, all of which are independent, at least in principle. Therefore, it is natural to try and find objects that are able to keep all of this information, without having to compute an infinite number of quantities. For this reason, we now turn to exploring 1- and 2-point probability density functions.

\subsection{Fixed-point probability distribution functions}

In the presence of a generic deviation from Gaussian statistics, involving both local and nonlocal terms, to assume that it is possible to capture all non-Gaussian information by looking at the single-point statistics of a field (\emph{i.e.}, correlations with all of the spatial coordinates at the same position) seems misguided, as the restriction to a single point is likely to mix local and nonlocal effects, making it difficult to disentangle them. However, if we restrict ourselves to local deviations from Gaussianity only, it is indeed possible to capture all such information.

In this subsection we write down explicit 1-point and 2-point PDFs for the curvature perturbation. In order to alleviate the discussion, we shall leave some of the details for App.~\ref{sec:2-point-det}.

\subsubsection{1-Point probability distribution function}

Now we set ourselves to derive the simplest distribution that can be obtained within this framework: a density function for the 1-point statistics. It is defined as the distribution $\PDF(\zeta;\x)$ that satisfies
\be
\langle \zeta^n(\x) \rangle = \int_{-\infty}^{\infty} \de\zeta \PDF(\zeta; \x) \zeta^n.
\ee
Given that we assume a homogeneous universe, $\PDF(\zeta;\x)$ cannot depend on $\x$. Thus, we write $\PDF(\zeta;\x) = \PDF(\zeta)$. From a functional perspective, it is given by
\be \label{1pt-condition}
\PDF(\bar{\zeta}) = \int D\zeta \, \PDF[\zeta] \, \delta(\zeta(\x) - \bar{\zeta}),
\ee
which may be evaluated in the same way as the partition function $Z[J]$ by writing down the Dirac delta as $\delta(\zeta(\x) - \bar{\zeta}) = \int_\gamma e^{i \gamma (\zeta(\x) - \bar{\zeta})}$ and noticing that what will be left with in the functional integral is exactly $Z[J(-\K) = \gamma e^{-i\K \cdot \x}] $. Then the remaining integral over $\gamma$ may be carried out by completing squares.

Thus, in the same spirit as the probability distribution functional we obtained earlier represents a first-order correction to Gaussian statistics, the 1-point probability density function may also be written as a slight departure from Gaussianity. Moreover, this density function resembles more closely the structure of $Z[J]$ than that of $\PDF[\zeta]$ because marginalising over all the other positions in presence of a finite range of wavelengths induces a filtering, which is manifest in $Z[J]$ but not so in $\PDF[\zeta]$. Given the various applications it may find in LSS or in primordial black hole formation~\cite{Panagopoulos:2019ail}, it is of interest to write it for an arbitrary window function. 

Smoothing the field and its variance as in Eq.~\eqref{z-filtered}, the resulting expression is
\be \label{one-point}
\PDF (\zeta_W) =\frac{1}{\sqrt{2\pi}\sigma_{W} } e^{-\frac{\zeta_W^2}{2\sigma_{W}^2}}  \left[ 1 + \Delta_W(\zeta_W) \right], 
\ee
where
\be \label{one-point-D}
 \Delta_W (\zeta_W) \equiv \int_0^{\infty} \!\! \de x \;  \frac{4\pi x^2 W(x)}{{\rm s}^2(x)} \! \int_{-\infty}^{\infty} \!\!\!\! \de \bar \zeta \,\, \frac{\exp \Big[{-\frac{ \left(\bar \zeta - {\rm b}(x) \zeta_W \right)^2}{2\sigma_{W}^2 (x)} } \Big] }{\sqrt{2\pi  } \sigma_{W} (x)}  \left( {\bar \zeta}  - \sigma_\zeta^2 \frac{\partial}{\partial {\bar \zeta}} \right) \! F \! \left({\bar \zeta}\right). 
\ee
Here we have written $W(x) = \int_{\K} e^{i \K \cdot \x} W(k)$, the position-space representation of the window function $W$, with $x = |\x|$, and we have also defined the (co)variances
\be \label{var-1-defs-t}
\sigma_{W}^2 \equiv \int_k W^2(k) P_\zeta(k)\;, \quad {\rm s}^2(x) \equiv \int_k e^{i\K \cdot \x} W(k) P_\zeta(k) \;, \quad
\sigma_{W}^2(x) \equiv \sigma_\zeta^2 -  {\rm b}^2(x) \sigma_{W}^2,
\ee
with a ``bias'' factor\footnote{This parameter is not the usual bias. If we think of the window as setting the scale of a tracer, we can define a bias as the ratio of field-tracer to the tracer-tracer correlation functions. ${\rm b}$ here is the ratio of the field-tracer to the 1-point tracer-tracer correlation functions.} given by
\be \label{b-1-defs-t}
{\rm b}(x) \equiv \frac{{\rm s}^2(x)}{\sigma_W^2}.
\ee

In words, this expression means that, given a range of modes defining the cutoffs of the theory, the perturbative correction to the PDF scans for the structure of the effective interaction at those scales through the action of $( {\bar \zeta} - \sigma_\zeta^2 \partial_{\bar \zeta} )$, and then filters it, through the Weierstrass transform, according to the difference between the variance and the correlation implied by incorporating window functions.
This observable can account for all the information contained within the local function $F$. This can be seen from the fact that all of the information concerning $F$ is stored within the $f_n$ coefficients, which can be retrieved by looking only at the 1-point statistics (\emph{c.f.} Eq.~\eqref{npt-position}). Indeed, an analysis aiming to constrain the departure from Gaussianity $\F_{\rm NG}$ by the means of the 1-point CMB temperature distribution has already been performed in~\cite{Chen:2018brw} .

Before passing to the 2-point functional, and in order to make contact with current literature, let us comment on the relation of this PDF to the Edgeworth representation. If we use the Hermite polynomial expansion of $F$, given in Eq.~\eqref{F-herm}, the effect of the Gaussian filtering of Eq.~\eqref{one-point-D} becomes transparent:
\be \label{W-effect}
\int_{-\infty}^{\infty} \!\!\!\! \de \bar \zeta \,\, \frac{\exp \Big[{-\frac{ \left(\bar \zeta - {\rm b}(x) \zeta_W \right)^2}{2\sigma_{W}^2 (x)} } \Big] }{\sqrt{2\pi  } \sigma_{W} (x)} \! F(\bar\zeta;\sigma_\zeta) = F\left( {\rm b}\zeta_W ;{\rm b}^2\sigma_W^2 \right) = \sum_{n=2}^\infty \frac{f_{n}}{n!} \,   {\rm He}_{n} \! \left( \frac{\zeta_W}{\sigma_W} \right) \frac{{\rm s}^{2n}(x)}{\sigma_W^{n}},
\ee
that is, it replaces the field variable with the biased one. Using the fact that the Weierstrass transform commutes with derivatives, the latter now being evaluated at the biased field with its corresponding variance, and that $(x-\sigma^2\partial_x){\rm He}_{n}(x/\sigma)=\sigma{\rm He}_{n+1}(x/\sigma)$, we can rewrite the NG deviation of the 1-point PDF as
\bea \label{Edgw}
\Delta_W (\zeta_W) &=&  \sum_{n=2}^\infty \frac{f_{n}}{n!} \,   {\rm He}_{n+1} \! \left( \frac{\zeta_W}{\sigma_W} \right) \frac1{\sigma_W^{n+1}} \int_0^{\infty}  \!\! \de x \;   4\pi x^2 W(x) {\rm s}^{2n}(x) \nonumber \\
&=&  \sum_{n=2}^\infty \frac{{\rm He}_{n+1} \! \left( \frac{\zeta_W}{\sigma_W} \right)}{(n+1)!} \, \frac{\langle \zeta_W^{n+1} \rangle_c}{\sigma_W^{n+1}},   
\eea
where $\langle \zeta_W^{n} \rangle_c$ is given by Eq.~\eqref{n-pt-f} integrated over momenta.
This is exactly the Edgeworth expansion of a non-Gaussian PDF truncated to first order in the couplings $f$, since we have restricted our derivation of the PDF~\eqref{one-point-D} to first order\footnote{In principle, there is nothing stopping us from computing the Edgeworth expansion to any order in the couplings; indeed, we can expand the exact PDF of Eq.~\eqref{exact-PDF} to any order in $F$. We choose, however, for simplicity to truncate the series to first order.} in $F$. 

Moreover, in this case it is possible to invert the Edgeworth expansion~\eqref{Edgw} in terms of the function $F$. Taking the $m$th Hermite moment of this expansion gives
\be
\frac{\langle \zeta_W^{m} \rangle_c}{\sigma_W^{m}} = \int_{-\infty}^{\infty} \de \zeta_W \frac{e^{-\frac{\zeta_W^2}{2 \sigma_W^2} }}{\sqrt{2\pi} \sigma_W } {\rm He}_m  \! \left( \frac{\zeta_W}{\sigma_W} \right) \Delta_W (\zeta_W),
\ee
and by replacing the cumulant $\langle \zeta_W^{m} \rangle_c$ in terms of $f_m$ we get
\be
f_n = \frac{ \sigma_W^{n+1}/(n+1) }{ \int_0^{\infty}  \!\! \de x \;   4\pi x^2 W(x) {\rm s}^{2n}(x)  }   \int_{-\infty}^{\infty} \de \zeta_W \frac{e^{-\frac{\zeta_W^2}{2 \sigma_W^2} }}{\sqrt{2\pi} \sigma_W } {\rm He}_{n+1}  \! \left( \frac{\zeta_W}{\sigma_W} \right) \Delta_W (\zeta_W),
\ee
which means that we have obtained the coefficients of the Hermite polynomial expansion~\eqref{F-herm} of $F$. Therefore, all one needs to write an explicit expression for $F$ is to perform the correlation integrals $ \int_0^{\infty}  \!\! \de x \;   4\pi x^2 W(x) {\rm s}^{2n}(x)$, and then sum back\footnote{The precise scaling with $n$ of the correlation integrals depends on the window function of choice $W$, and thus it is not always possible to give the result for $F$ in a closed form.} its Hermite expansion~\eqref{F-herm}. This allows for a direct test of NG in a given dataset and provides a way of constraining the primordial quantities involved in its generation, in a manner that is identical to what has been previously done in~\cite{Chen:2018brw}.

However, this type of analysis integrates out information about the momentum dependence of the correlation functions, which can be used to place tighter constraints on the parameters of a given model. We now proceed to explore the two-point PDFs for the filtered primordial perturbations, which do contain such information.

\subsubsection{2-Point probability distribution function}

Now we would like to write down an observable able to account for the non-Gaussian statistics, without integrating out the information about correlations in the sky at different scales. Therefore it cannot be a single-point PDF. Thus, we try to do the next least complicated thing: a 2-point PDF $ \PDF (\zeta_1, \zeta_2; \x_1, \x_2)$. This function satisfies
\be\label{2-pr-corr}
\langle \zeta_W^n(\x_1) \zeta_W^m(\x_2) \rangle = \int \de\zeta_1 \de\zeta_2 \; \PDF_W (\zeta_1, \zeta_2; |\x_1-\x_2|) \zeta_1^n \zeta_2^m,
\ee
where we have written $ \PDF (\zeta_1, \zeta_2; \x_1, \x_2) = \PDF (\zeta_1, \zeta_2; |\x_1-\x_2|)$ because we are assuming our universe to be statistically homogeneous. One way to obtain such an object is by conditioning in two points in a manner analogous to Eq.~\eqref{1pt-condition},
\be \label{2pt-condition}
\PDF(\zeta_1, \zeta_2; |\x_1 - \x_2|) = \int D\zeta \, \PDF[\zeta] \, \delta(\zeta(\x_1) - \zeta_1) \delta(\zeta(\x_2) - \zeta_2) ,
\ee
where again, the final result can only depend on the spatial coordinates through the distance between the two positions $\x_1$ and $\x_2$. With this in mind, let us define the scalar variables
\be
r\equiv|\x_1 - \x_2|,\quad r_1\equiv|\x - \x_1|,\quad r_2\equiv|\x - \x_2|.
\ee

Then, by inspecting the $n$-point functions of the theory~\eqref{npt-position} evaluated at the two points of interest, we obtain a similar expression as the 1-point PDF, but with two points defining the filtering instead of one:
\be \label{two-point}
\begin{split}
\PDF_W (\zeta_1,\zeta_2,r) = & \; \PDF_{G,W}(\zeta_1,\zeta_2,r)   \Bigg[ 1 - \int_\x \int_{-\infty}^{\infty} \!\!\!\! \de \bar \zeta   \frac{\exp \Big[{-\frac{ \left(\bar \zeta - \zeta_W(r,r_1,r_2) \right)^2}{2\sigma_W^2 (r,r_1,r_2)} } \Big] }{\sqrt{2\pi  } \sigma_W (r,r_1,r_2)}  \\ &  \times \left\{  \frac{W(r_1)}{s^2(r_1)} \!  \left( G_{11} \frac{\partial}{\partial {\bar \zeta}} - G_{12} \right)      +  \frac{W(r_2)}{s^2(r_2)} \!  \left( G_{21} \frac{\partial}{\partial {\bar \zeta}} - G_{22} \right) \right\} F  \! \left({\bar \zeta}\right)  \Bigg],
\end{split}
\ee
where $\PDF_{G,W}(\zeta_1,\zeta_2,r)$ is the bivariate Gaussian measure, with a covariance matrix given by the $2\times2$ bottom right block of $\pmb{\Sigma}$, defined below in Eq.~\eqref{cov}.
Let us go through this expression: 
the first thing to notice is the presence of two points, $\x_1$ and $\x_2$, defining a filtering through the same function as in the 1-point case. The second important aspect is that now the Gaussian that is convoluted with $F$ has a different mean and variance. However, they emerge in the same manner as $\sigma_W^2(x)$ and $\zeta_W(x)$ emerge in the 1-point case: $\sigma^2_W(r,r_1,r_2)$ and $\zeta_W(r,r_1,r_2)$ are the variance and mean of $\bar \zeta$ after conditioning on the values of $(\zeta_1, \zeta_2)$, starting from a joint Gaussian distribution for $({\bar \zeta}, \zeta_1, \zeta_2)$ with covariance matrix
\be \label{cov}
  \pmb{\Sigma}=
  \left[ {\begin{array}{ccc}
   \sigma_\zeta^2 & s^2(r_1) & s^2(r_2) \\
   s^2(r_1) & \sigma_W^2 & \sigma_{W,{\rm ext}}^2(r)\\
   s^2(r_2) & \sigma_{W,{\rm ext}}^2(r) & \sigma_W^2 \\
  \end{array} } \right],
\ee
where we have written the covariance between the two externally chosen points $\x_1$ and $\x_2$ as
\be
\sigma_{W,{\rm ext}}^2(r) = \int_k e^{i\K \cdot (\x_1 - \x_2)} W^2(k) P_\zeta(k).
\ee
The functions $G_{ij}$ also appear in a similar way: $G_{i1}$ and $G_{i2}$ are ``rotated'' versions of $\sigma_\zeta^2$ and $\bar \zeta$, involving combinations of the free theory covariances that make the overall expression reduce to that of the 1-point PDF as $\x_1 \to \x_2$. Their precise definitions are listed in App.~\ref{sec:2-point-det}. In there, we delineate how to obtain the 2-point PDF in an alternative way: we start from correlators of the type $\langle \zeta_W^n(\x_1) \zeta_W^m(\x_2) \rangle$ and we deduce the function from which they emanate, corresponding to~\eqref{2-pr-corr}.

This PDF contains all the information of the free theory, as having two points allows to scan over all the range of distances in the sky, thus probing, among others, the 2-point correlation function completely, which is the defining object of a Gaussian theory. Even though Eq.~\eqref{two-point} has its non-Gaussian features encoded in a perhaps more complicated fashion than its 1-point counterpart~\eqref{one-point}, both contain the same information about the underlying function $\F_{\rm NG}$. Indeed, one can obtain Eq.~\eqref{one-point} by integrating over one of the field variables in Eq.~\eqref{two-point}. However, observationally, it might be more efficient to have information on the scale, since then we can, for example, disentangle different momentum shapes of correlation functions.

\subsection{Linear transfer functions and their limitations} \label{sec:scnd-order-pt}

The above discussion sets a clear framework to study various cosmological observables in the late-time universe with linear perturbation theory. In particular, we can consider endowing the window function $W(k)$ with an additional argument, thus defining the \textit{transfer function} $T({\bf a},\K)$, to obtain an observable $A({\bf a})$, where $\bf a$ is a label for the new quantity/field that is linearly related to $\zeta$ as
\be
A({\bf a}) = \int_\K T({\bf a},\K) \zeta(\K),
\ee
up to first order in cosmological perturbation theory. In the case of CMB and LSS, the temperature fluctuations map
\be
\Theta(\n) = \int_\K T(\n,\K) \zeta(\K),
\ee
and the matter density contrast
\be
\delta(\x) = \frac{3}{5} \int_\K e^{i \K \cdot \x} \alpha(k) \zeta(\K),
\ee
are well described by linear cosmological perturbation theory as long as the distance/momentum scales are sufficiently large/small so that late-time nonlinearities do not affect them. This is indeed the case for the CMB scales $10^{-4}\;{\rm Mpc}^{-1} \leq k \leq 0.3\; {\rm Mpc}^{-1}$ {\sc Planck} observed~\cite{Akrami:2018odb,Akrami:2019bkn}, and for the large-scale structure of our universe provided that $k \lesssim 0.1 \; {\rm Mpc}^{-1}$ and the redshift $z$ satisfy $z \gtrsim 2$~\cite{Sefusatti:2007ih} so that gravitational nonlinear effects remain suppressed. Sections~\ref{sec:theta-stat} and~\ref{sec:delta-stat} delve into statistical objects and probes that can be used to test for imprints of non-Gaussianity in the CMB and LSS, respectively, in this way.

However, even though for {\sc Planck} and {\sc Wmap} linear perturbation theory is indeed accurate enough, as the errorbars are expected to keep shrinking, this might not be the case for future surveys~~\cite{Smith:2009jr,Seljak:1998nu,Goldberg:1999xm,Verde:2002mu,Babich:2008uw,Pitrou:2008ak,Bartolo:2008sg,Senatore:2008vi,Senatore:2008wk,Nitta:2009jp,Bartolo:2004if,Bartolo:2006cu,Bartolo:2006fj,Bartolo:2005fp,Creminelli:2004pv}.
At this point, it is thus appropriate to clarify that our subsequent discussion, as it stands, is valid for perturbation theory with linear transfer functions, and as such, it will have some limitations inherited from this approach. For example, we will not take into account the fact that the gravitational potentials differ at second order, nor will we use the nonlinear transfer function corresponding to such higher order terms~\cite{Bartolo:2005kv,Bartolo:2006cu,Bartolo:2006fj,Bartolo:2005fp,Boubekeur:2009uk}. We also do not discuss nonlocal effects like the ISW to second order, which can be included in the window function via a convolution in time. That would complicate the formulas because it would force us to include additional integrals over the history of our universe to preserve the generality of the approach (and not only over a single spatial slice), which could in turn reduce the usefulness of defining transfer functions in our setup, as including integrals over time might often be equivalent to following the dynamical evolution of the perturbations up to some fixed order in time-dependent perturbation theory. The alternative would be to carry out a traditional perturbative computation, without resorting to a probability distribution function in the process.  
There remains, hence, the question of what the information extracted from measurements means for our understanding of the primordial universe: since the late-time nonlinear evolution is only treated perturbatively, at some order one might lose primordial information in the local ansatz because nonlinearities in the later evolution of the universe will modify the observable consequences of the primordial $\F_{\rm NG}$, and consequently, the primordial PDF we would infer for $\zeta$.

We note, however, that if the experimental precision demands to have control of the cosmological perturbations up to second order, one can intent to proceed further and include such effects by writing the full temperature field (or the matter density contrast $\delta$) as $\Theta=\Theta^{(1)} + \Theta^{(2)}+\cdots$. One could then induce a PDF for the fully nonlinear field taking into account both primordial and secondary NG to any desirable order\footnote{By doing so, one would have to consider the field and its gradient as independent  Gaussian variables, since the latter contributes to the second order curvature perturbation~\cite{Bartolo:2005kv}.}, allowing for generic nonlocal mappings $\F_{\rm NG}$. In other words, one can parametrise the generalised ansatz as in Eq.~\eqref{general-ansatz-restricted}, in the manner of an effective object that adequately describes the late-time observables, where the expansion coefficients account for nonlinearities in both the initial conditions and due to gravitational evolution~\cite{Bartolo:2005fp}. In~\cite{Nitta:2009jp,Creminelli:2004pv} it was shown that second order effects provide for corrections to the skewness (albeit with distinguishable signals from the primordial component) around the sensitivity of future CMB surveys like CMB-S4, and one might, arguably, worry that this could also be the case for higher order correlators like the trispectrum; it is, however, conceivable that the secondary effects would keep diminishing as one considers higher order statistics.
A way out could thus be a case-by-case evaluation of the secondary effects in higher order $n$-point functions~\cite{Bartolo:2005fp}, or a full characterisation of the secondary effects at the level of the PDF, via, \emph{e.g.}, simulations.

Even though, admittedly, we do not offer any crucial insight into the problem of disentangling primordial from late-time NG, our results set the stage for further study of NG signatures on PDF estimators, especially targeting models beyond single-field inflation (\emph{c.f.} App.~\ref{sec:axion}), where primordial local NG is enhanced compared to other shapes of NG and secondary effects.

\setcounter{equation}{0}

\section{$\Theta$ Statistics} \label{sec:theta-stat}

Having set the stage for studying late-time cosmological perturbations, we can now turn to discussing our specific scenarios of interest. To begin with, it is of particular relevance to write down testable quantities that we can obtain by looking at the primordial information that can be stored in spherical shells on the sky, such as the CMB. Namely, both the probability distribution functional $\PDF[\zeta]$ and its associated partition function $Z[J]$ may be projected onto the celestial sphere to yield distributions of the temperature fluctuations $\delta T(\hat{n})$.

Consider a generic linear transfer function $T(\K,\n)$ from the primordial perturbations $\zeta$ to an observable defined on the sphere $\Theta(\n)$ such that
\be
\Theta(\n) = \int_\K T(\n,\K) \zeta(\K).
\ee
Of course, we are interested in the specific case where $\Theta (\hat{n}) = \delta T(\hat{n}) = (T(\hat{n}) - T_0)/T_0$ (where $T(\hat{n})$ is the CMB temperature measured in a given direction $\hat n$, and $T_0$ its average over all directions), however, the present discussion is also valid for more general observables that depend on the line of sight $\hat n$.
Then, if we take $\Sigma(\n,\n')$ to be the covariance matrix defining the observable's correlations between different directions in the sky $(\n,\n')$, \emph{i.e.}, 
\be
\langle \Theta(\n) \Theta(\n') \rangle = \Sigma(\n,\n') ,
\ee
and $\Sigma^{-1}(\n,\n')$ as its inverse matrix, we find that the probability distribution functional for $\Theta$ is given by
\be
\begin{split}
\PDF[\Theta] = \mathcal{N}_{\Theta} e^{-\frac{1}{2} \int_{\n} \int_{\n'} \Theta(\n) \Sigma^{-1}(\n,\n') \Theta(\n') } & \left[ 1 - \int_\x  {\rm Ker_1}(\x) \frac{\partial F_\Theta}{\partial \zeta }(\zeta_{\Theta}(\x);\x) \right. \\ & \,\,\,\,\,\,\, \left. + \int_\x {\rm Ker_2}(\Theta; \x) F_\Theta(\zeta_\Theta(\x);\x) \right],
\end{split}
\ee
where now ${\rm Ker_1}(\x)$, ${\rm Ker_2}(\Theta;\x)$, $\zeta_{\Theta}(\x)$, and $F_\Theta(\zeta;\x)$ depend implicitly on the transfer function and, when noted, the variable of interest $\Theta(\n)$. All of these quantities may be intuitively understood as the result of projecting a field defined on three spatial dimensions over a two-dimensional spherical surface. For instance, $\zeta_{\Theta}(\x)$ is given by
\be
\zeta_{\Theta}(\x) \equiv \int_\K \int_{\n} \int_{\n'} T(\n,\K) e^{i\K \cdot \x} P_\zeta(k) \Sigma^{-1}(\n,\n') \Theta(\n')  ,
\ee
which basically amounts to saying: take the original statistics of your theory, \emph{i.e.}, $e^{i\K \cdot \x} P_\zeta(k)$, project them onto the sphere by applying $T(\n,\K)$ and then correlate it with the field of interest $\Theta(\n')$ by means of its inverse covariance matrix $\Sigma^{-1}(\n,\n')$. The integration kernels have similar definitions:
\bea
{\rm Ker_1}(\x) &\equiv & \int_\K \int_{\K'} \int_{\n} \int_{\n'} T(\n,\K) e^{i\K \cdot \x} P_\zeta(k) \Sigma^{-1}(\n,\n') T(\n',\K') e^{i\K' \cdot \x} , \\
{\rm Ker_2}(\Theta; \x) &\equiv& \int_\K \int_{\n} \int_{\n'} T(\n,\K) e^{i\K \cdot \x} \Sigma^{-1}(\n,\n') \Theta(\n').
\eea
However, the function $F_\Theta$ is a slightly different object than before. As a result of the projection, it acquires a spatial dependence, whose exact nature in terms of the primordial departure from Gaussianity $F$ is given by the Weierstrass transform:
\be
F_\Theta(\zeta; \x) \equiv \exp \left[ \frac{\sigma_\zeta^2 - \sigma_{\zeta \Theta}^2(\x)}{2} \frac{\partial^2 }{\partial \zeta^2} \right] F(\zeta)  = \int_{\bar \zeta} \frac{\exp \Big[{-\frac{ \left(\bar \zeta - \zeta \right)^2}{2(\sigma_\zeta^2 - \sigma_{\zeta \Theta}^2(\x))} } \Big] }{\sqrt{2\pi (\sigma_\zeta^2 - \sigma_{\zeta \Theta}^2(\x)) } } F(\bar \zeta),
\ee
where $\sigma_{\zeta \Theta}^2(\x)$ may be understood as a position-dependent effective variance of $\zeta$, modified by projection effects:
\be
\sigma_{\zeta \Theta}^2(\x) \equiv \int_\K \int_{\K'} \int_{\n} \int_{\n'} T(\n,\K) e^{i\K \cdot \x} P_\zeta(k) \Sigma^{-1}(\n,\n') T(\n',\K') e^{i\K' \cdot \x} P_\zeta(k').
\ee
In all of the above, we have written $\Sigma(\n,\n')$ as a general function of the direction on the sphere. However, if we take into account that our universe is homogeneous, it must be possible to write it as a function of the angle between the two vectors, or equivalently, in terms of their scalar product $\n \cdot \n'$. We will not overemphasize this in what follows, as the notation we deem natural to treat $\n$ and $\n'$ is with them as separate directions, because they will have to be multiplied with another vector, the integration variable $\x$ that will appear in the NG kernel that modifies the PDF, which makes using $\n \cdot \n'$ notationally heavier than just using $(\n,\n')$.

Now we have to find ways of using this. One option would be to compare how likely is our present-day CMB given a certain primordial deviation from Gaussianity of the local type $F$ by using Bayesian statistics. However, in order for this to be useful at its maximum capacity, it is likely that one would first have to establish a definitive imprint of primordial NG and be forced to introduce extra parameters into the effective description because more often than not a model comparison will favour the one with less parameters. Therefore, we will turn to the observable we outlined at the end of Sec.~\ref{sec:zeta-stat}, which, to our knowledge, has been largely unexplored in this context\footnote{The 2-point PDF has been used in LSS \emph{count-in-cells} statistics~\cite{Codis:2016dyz}.} and may offer valuable constraints on the nature of primordial NG: a 2-point PDF. We present this result after briefly discussing the 1-point statistics.

\subsection{1-Point temperature distribution function}

Given the discussion we have followed so far, it is now straightforward to reproduce the results of~\cite{Chen:2018brw} and compute the 1-point PDF of the temperature map. The only difference is that the transfer functions map the fluctuations to the celestial sphere instead of the three-dimensional universe, but this does not prevent us from finding the distribution that generates the correlations. In analogy to~\eqref{one-point}, the result is given by
\be \label{one-point-T}
\PDF (\Theta) =\frac{1}{\sqrt{2\pi}\sigma_{\Theta} } e^{-\frac{\zeta_\Theta^2}{2\sigma_{\Theta}^2}}  \left[ 1 + \Delta_\Theta(\Theta) \right], 
\ee
where
\be \label{one-point-T-D}
 \Delta_\Theta (\Theta) \equiv \int_\x  \frac{W_\Theta(\x,\n)}{{\rm s}_\Theta^2(\x,\n)} \! \int_{-\infty}^{\infty} \!\!\!\! \de \bar \zeta \,\, \frac{\exp \Big[{-\frac{ \left(\bar \zeta - {\rm b}_\Theta(\x,\n) \Theta \right)^2}{2\sigma_{\Theta}^2 (\x,\n)} } \Big] }{\sqrt{2\pi  } \sigma_{\Theta} (\x,\n)}  \left( {\bar \zeta}  - \sigma_\zeta^2 \frac{\partial}{\partial {\bar \zeta}} \right) \! F \! \left({\bar \zeta}\right),
\ee
with the corresponding definitions of the window function
\be
W_\Theta(\x,\n) = \int_{\K} e^{i \K \cdot \x} T(\K,\n),
\ee
and that of the variances
\bea
\sigma_{\Theta}^2 \equiv \int_k T^2(\K,\n) P_\zeta(k) \, , \\
 {\rm s}_\Theta^2(\x,\n) \equiv \int_k e^{i\K \cdot \x} T(\K,\n) P_\zeta(k) \, , \\
\sigma_{\Theta}^2(\x,\n) \equiv \sigma_\zeta^2 -  {\rm b}_\Theta^2(\x,\n) \sigma_{\Theta}^2 \, ,
\eea
with $b_\Theta$ given by
\be 
{\rm b}_\Theta(\x,\n) \equiv \frac{{\rm s}_\Theta^2(\x,\n)}{\sigma_\Theta^2}.
\ee
The final result is independent of the angular direction $\n$ because of homogeneity. In practice this appears because the form of the transfer functions $T(\K,\n)$ will imply that the only dependence on $\n$ will be through scalar products with $\x$ or $\K$, which as integration variables span the whole space, and thus they render all directions statistically homogeneous.

In Ref.~\cite{Chen:2018brw}, an analysis of the CMB statistics was carried out to search for non-Gaussian signals by estimating the PDF~\eqref{one-point-T} directly from the 2015 Planck dataset~\cite{Adam:2015rua}, with an inconclusive result: if primordial non-Gaussianity of this type is present in the CMB, its statistical imprints are smaller than the intrinsic noise to which the 1-point statistics are subject to. With this and the forthcoming surveys in mind, we now proceed to outline another statistical object whose reconstruction from observations shows promise to have a lower intrinsic noise but still preserve all of the information contained within the 1-point PDF: a 2-point PDF. 

\subsection{2-Point temperature distribution function}

It is interesting to write down expressions for quantities that are not typically used directly when characterising cosmological datasets. For instance, one rarely bothers to write down the full 2-point PDF for the CMB temperature map, as all of its information (in the Gaussian case) is already specified through the power spectrum. However, the scale dependence of this PDF may be a useful tool to search for non-Gaussianities. The result, analogously to what we had in the simpler case of a spatial window function, is
\be \label{two-point-th}
\begin{split}
\PDF_\Theta (\Theta_1,\Theta_2,\n_1,&\n_2) = \; \PDF_{G,W}(\Theta_1,\Theta_2,\n_1,\n_2)   \Bigg[ 1 - \int_\x \int_{-\infty}^{\infty} \!\!\!\! \de \bar \zeta   \frac{\exp \Big[{-\frac{ \left(\bar \zeta - \zeta_\Theta(\x,\n_1,\n_2) \right)^2}{2\sigma_\Theta^2 (\x,\n_1,\n_2)} } \Big] }{\sqrt{2\pi  } \sigma_\Theta (\x,\n_1,\n_2)}  \\ & \;\times   \left\{  \frac{W_\Theta(\x, \n_1)}{s_\Theta^2(\x , \n_1)} \!  \left( G_{11}^{\Theta} \frac{\partial}{\partial {\bar \zeta}} - G_{12}^{\Theta} \right)      +  \frac{W_\Theta(\x, \n_2)}{s_\Theta^2(\x,  \n_2)} \!  \left( G_{21}^{\Theta} \frac{\partial}{\partial {\bar \zeta}} - G_{22}^{\Theta} \right) \right\} F  \! \left({\bar \zeta}\right)  \Bigg].
\end{split}
\ee
As before, let us go through this expression: the first thing to notice is that the two filtering points $\x_1$ and $\x_2$ in~\eqref{two-point} are now replaced by two directions in the sky, $\n_1$ and $\n_2$. Secondly, the Gaussian that is convoluted with $F$ now has a different mean and variance, obtained by conditioning the joint multivariate Gaussian PDF on the values of $\Theta_1$ and $\Theta_2$. 
The starting point to this is a joint Gaussian distribution for $({\bar \zeta}, \Theta_1, \Theta_2)$ with covariance matrix
\be \label{cov-th}
  \pmb{\Sigma}=
  \left[ {\begin{array}{ccc}
   \sigma_\zeta^2 & s_\Theta^2(\x, \n_1) & s_\Theta^2(\x , \n_2) \\
   s_\Theta^2(\x, \n_1) & \sigma_\Theta^2 & \sigma_{\Theta,{\rm ext}}^2(\n_1, \n_2)\\
   s_\Theta^2(\x , \n_2) & \sigma_{\Theta,{\rm ext}}^2(\n_1, \n_2) & \sigma_\Theta^2 \\
  \end{array} } \right],
\ee
where we have written the covariance between the two externally chosen points $\n_1$ and $\n_2$ as
\be
\sigma_{\Theta,{\rm ext}}^2(\n_1, \n_2) = \int_k T(\n_1,\K) T(\n_2,\K)  P_\zeta(k) = \Sigma(\n_1, \n_2),
\ee
and $\sigma_\Theta^2 = \sigma_{\Theta,{\rm ext}}^2(\n, \n)$, which is independent of the direction $\n$. The precise definitions of all the additional functions involved in this section are given in App.~\ref{sec:2-point-det}. It is worth mentioning, as a reminder to the reader, that, as in the 2-point PDF for curvature fluctuations~\eqref{two-point}, both~\eqref{two-point-th} (besides from the temperature variables) and $\Sigma(\n_1,\n_2)$ depend only on the angular distance between $\n_1$ and $\n_2$, or equivalently, on $\n_1 \cdot \n_2$, and that this is a consequence of our universe's homogeneity.

    From this function, \emph{i.e.}, from the 2-point PDF~\eqref{two-point-th}, it is possible to obtain refined constraints on the local ansatz. Given a dataset, one can construct the 2-point PDF as follows: divide the angular distance into $N$ bins of width $\delta\vartheta$ and the temperature in $M \times M$ bins of size $\delta\Theta \times \delta\Theta$, as in the 1-point PDF but now with two axes for the temperature field. Now, for each bin associated to a given angular distance $\vartheta_n = n \cdot \delta\vartheta$, and for each value of $(i,j)$, count how many pairs of pixels separated by that angular distance $\vartheta_n$ have the values $(\Theta_{[i]}, \Theta_{[j]})$ for the temperature in their respective positions. This process would generate $N$ two-dimensional histograms, with two temperature axes, which we label by $(\Theta_1,\Theta_2)$, whose value at coordinate $({\Theta_1}_{[i]}, {\Theta_2}_{[j]})$ would give the number of pairs of pixels with temperatures in the $(i,j)$th bin, separated by angular distance in the $n$th bin. A $1/2$ symmetry factor must be included in the number counts for temperature bins with $i\neq j$, as the bin $({\Theta_1}_{[i]}, {\Theta_2}_{[j]})$ is equivalent to $({\Theta_1}_{[j]}, {\Theta_2}_{[i]})$.
    
    How does this give refined constraints on the local ansatz? Let us appreciate that this set of PDFs contains information on the scale, or more concretely, on the temperature power spectrum and of its expansion in spherical harmonics (the standard $C_\ell$s) through the 2-point correlation $\Sigma(\n_1,\n_2)$. If NG is absent, then at each value of the angular distance the 2-point PDF will be a 2-variable Gaussian probability density with variances $\sigma_\Theta^2$ and covariance $\sigma_{\Theta, {\rm ext} }^2(\n_1,\n_2) = \Sigma(\n_1,\n_2)$. Then, in the presence of NG, each 2-point PDF will undergo a NG deviation induced by the same primordial mechanism $F$, but for each angular separation this deviation will be experienced differently because the covariance matrix implied by the Gaussian part is different. This means that for each angular distance, the kernel that acts upon $F$ in~\eqref{two-point-th} gives a different deviation from Gaussianity, and therefore, each of the 2-point PDFs gives an independent estimator on the primordial NG field. For local NG, all of the $N$ 2-point PDFs at different angular separations in the sky should give consistent\footnote{That is, within the experiment's theoretical and systematic uncertainties.} constraints/estimations of $F$. Conversely, if NG is measured and it does not adjust to the statistics implied by~\eqref{two-point-th} at different angular scales, then purely local NG would be ruled out. Therefore, looking towards possible future directions to be explored, this type of object (a set of 2-point PDFs) shows promise to disentangle different shapes of NG, such as equilateral or orthogonal templates, in particular, from the local ansatz.

  In order to search for non-Gaussianity within a 2-point PDF, many approaches are possible. Given a model, \emph{i.e.}, an explicit expression for $F$, and using it as a template with few adjustable parameters is usually the method that will give the best constraints. In the spirit of Eq.~\eqref{F-herm}, however, another one is worth mentioning: one can use bivariate Hermite polynomials on the temperature variables $(\Theta_1, \Theta_2)$, so as to express the PDF in terms of a bivariate Edgeworth expansion~\cite{Sellentin:2017aii,WITHERS2000165}. One then looks for any statistically significant nonzero coefficient in the expansion, in analogy to what was done for the 1-point case in~\cite{Chen:2018brw}. This way, the existence of NG can be tested as a yes/no question, as any nonzero coefficient in an Edgeworth expansion implies a non-Gaussian distribution. This may be particularly useful when searching for NG in the next generation CMB surveys~\cite{Abazajian:2016yjj}.

\setcounter{equation}{0}
\section{$\delta$ Statistics} \label{sec:delta-stat}

Upcoming cosmological surveys will focus on the statistics of LSS, promising to bring precision cosmology to a new era. Indeed, the proliferation of observed modes due to the three-dimensional probe offered by LSS will enhance our statistics giving us invaluable information about the fundamental aspects of the early/late universe. 

At small scales several sources of nonlinearity induce NG, like gravitational interactions and galaxy bias, obscuring the primordial contribution to the statistics. However, for relatively long modes, $k\lesssim 0.1\;$Mpc$^{-1}$ and higher redshifts~\cite{Sefusatti:2007ih}, linear perturbation theory can be trusted, which makes it easier to identify primordial signatures. Perturbative techniques pushing our analytic control towards smaller, weakly nonlinear scales include several schemes like SPT~\cite{Bernardeau:2001qr} and more recently EFTofLSS~\cite{Carrasco:2012cv} and TSPT~\cite{Blas:2015qsi,Blas:2016sfa,Vasudevan:2019ewf}, which are set within a hydrodynamics framework, while going even further requires full Boltzmann solvers via N-body simulations. In this work, we will focus on the purely linear regime, leaving weakly nonlinear evolution with NG initial conditions~\cite{Vasudevan:2019ewf} for future study.

The main probes of non-Gaussianity are the bispectrum and/or trispectrum, number counts and bias. The spectra retain information about the shape of the 3- and 4-point functions in momentum space, which can be linked to the mechanism responsible for generating NG.
Number counts probe directly the 1-point PDF, which even though loses the shape information, it serves as a complementary and equally powerful estimator of NG~\cite{Matarrese:2000iz, LoVerde:2007ri, Afshordi:2008ru, Valageas:2009vn, Musso:2011ck, Uhlemann:2017tex, Ivanov:2018lcg}. Finally, the halo bias serves as a third independent channel, which can give clear enhanced signatures of local NG at large scales.

In this section, we wish to track how primordial NG, in the form of the generic local ansatz~\eqref{general-ansatz-restricted}, gets transmitted to the matter field in the linear regime. We extend our result, the non-Gaussian PDF of curvature fluctuations, in two directions: 1) we deduce a PDF for the matter density contrast $\delta$, and hence, a halo mass distribution; 2) we compute the effect of the generalised local ansatz on the halo bias. These are complementary probes of the non-Gaussian initial condition via cluster number counts and power spectra, respectively, which should be accessible by surveys such as {\sc Lsst}, {\sc Euclid}, {\sc Sphere}x and {\sc Ska}.

\subsection{Halo mass function} \label{sec:halo}
The matter overdensity $\delta(\x) = \delta \rho(\x)/\bar{\rho}$, with $\bar{\rho} =\Omega_m\rho_{\rm cr}$, is related to the primordial Newtonian potential, $\Phi=\frac{3}{5} \zeta$, as
\be \label{poisson}
\delta(\K) = \alpha(k) \Phi(\K), \qquad {\rm with} \qquad \alpha(k)=\frac{2r_H^2k^2 {\mathcal T} (k)D(z)}{3\Omega_m},
\ee
where $D(z)$ is the linear growth rate, $r_H$ the current Hubble radius and ${\mathcal T}(k)$ the transfer function~\cite{Eisenstein:1997ik}. 
We smooth the density field over a radius $R_M=\left(3M/4\pi\bar{\rho}\right)^{1/3}$ as in Eq.~\eqref{z-filtered}, 
\be \label{d-filtered}
\delta_W (\x) = \int_\K \int_\y e^{i{\K} \cdot (\x - \y)} W_M(k) \delta(\K),
\ee
using a top-hat filter $W_M(x)=V_M^{-1}{\rm H}(R_M-x)$, with H the Heaviside function and $V_M$ the volume enclosing a mass M. 
The probability distribution for the smoothed overdensity, $\PDF (\delta_W)$, is then given by the 1-point PDF of Eq.~\eqref{one-point} upon the replacements
\be \label{z-to-d}
\zeta\to\delta \quad \text{and} \quad W(k)=\frac35W_M(k)\alpha(k).
\ee
Now, in principle, having the matter distribution function one can compute the halo number density, that is, the number density of halos of mass between $M$ and $M+\de M$ at redshift $z$, and the observables derived from it. One way to do this is via the Press-Schechter (PS) scheme~\cite{Press:1973iz} extended to the NG case~\cite{Lucchin:1987yv,Chiu:1997xb}. Let
\be  \label{mu}
\mu^>(M,z)=\int_{\nu_c(z)}^\infty \de \nu \; \PDF(\nu),
\ee
(with $\nu\equiv\delta/\sigma_W$) be the tail distribution above some threshold value $\delta_c(z)$. Then, one can assert that the total fraction of mass collapsed into bound structures will be proportional to this cumulative PDF:
\be
\frac{1}{\bar\rho}  \int_{M}^{\infty} \!\! \de m \frac{\de n}{\de m} \, m  \propto\mu^>(M),
\ee
where $\de n(M)$ is the number density of halos with masses in the range $M$ and $M+\de M$. The PS mass function then reads\footnote{The fudge factor $2$ corrects for the \emph{cloud-in-cloud} problem, that is, a collapsed object of mass $M_1$ tracing a volume $V_1$ can be missed by the PS function if it is part of an underdense region of larger volume $V_2$. The Gaussian value $2$ is a good approximation in case of a small NG deformation~\cite{Matarrese:2000iz}.}
\be  \label{F_PS}
\frac{\de n_{\rm PS} }{\de M}(M,z)=-2\frac{\bar\rho}{M} \frac{\de \mu^>}{\de M}(M,z).
\ee
The collapse threshold, through which the $z$ dependence arises, is taken to be given by the spherical model as $\delta_c(z)\simeq 1.686\;D(0)/D(z)$. The Gaussian PS function can be evaluated exactly by replacing $\PDF\to\PDF_{\rm G}$ in Eq.~\eqref{mu}:
\be  \label{F_PS_G}
\frac{\de n_{\rm PS}}{\de M}\Big|_{\rm G}(M,z)
= 2\frac{\bar\rho}M\frac{e^{-\frac{\nu_c^2(z)}{2}}}{\sqrt{2\pi}}  \frac{\de \nu_c(z)}{\de M}.
\ee
For the NG case, the PS mass function is easy to compute from Eq.~\eqref{one-point}:
\be \label{mass-function-PS-F}
\begin{split}
\frac{\de n_{\rm PS}}{\de M}(M,z) = \frac{\de n_{\rm PS}}{\de M}\Big|_{\rm G} (M,z)\left[1-\frac1{\delta_c}\left(1-\nu_c^2+\frac1{(\ln\nu_c)'} \frac{\de }{\de M}\right)\int_\x W(x)F\left( {\rm b}\delta_c;{\rm b}^2\sigma_W^2 \right) \right] ,
\end{split}
\ee
where a prime stands for the derivative with respect to the mass, while $F\left( {\rm b}\delta_c;{\rm b}^2\sigma_W^2 \right)$ is the Weierstrass transform of the local ansatz [see Eq.~\eqref{W-effect}] evaluated at the threshold $\delta_c(z)$ (note that here, $W(x)$ is the Fourier transform of the window function written in Eq.~\eqref{z-to-d}). 
Upon using the Hermite expansion~\eqref{F-herm} of the function $F$, we may obtain a series representation of the mass function~\eqref{mass-function-PS-F}:
\be \label{mass-function-PS-H}
\begin{split}
\frac{\de n_{\rm PS}}{\de M}(M,z) = \frac{\de n_{\rm PS}}{\de M}\Big|_{\rm G}(M,z)  \left[1+\Delta\left(\nu_c(z)\right)  - \frac{1}{\de\nu_c(z)/\de M} \sum_{n=2}^\infty \frac{{\rm He}_{n}\left(\nu_c(z)\right)}{(n+1)!} \kappa_{n+1}' \right] ,
\end{split}
\ee
where we have defined the reduced cumulants $\kappa_{n}(M)\equiv \frac{\langle \delta_W^{n} \rangle_c}{\sigma_W^{n}}$. 
Equation~\eqref{mass-function-PS-F} offers a generalisation of the $f_{\rm NL}$, $g_{\rm NL}$ truncation\footnote{For example, when truncated to $n=2$, the expansion~\eqref{mass-function-PS-H} agrees with Eq.~(4.19) of Ref.~\cite{LoVerde:2007ri}.} of the local ansatz (see \emph{e.g.} Refs.~\cite{LoVerde:2007ri,LoVerde:2011iz}) to arbitrary functions $F$. From its moment expansion~\eqref{mass-function-PS-H}, we can see that positive moments of the PDF ($\kappa_n,\kappa'_n>0$) lead to overabundance of collapsed objects at the high mass end, where $\nu,\nu'\gg1$, as long as $(\ln \kappa_{n+1})'<\nu\nu'$, which is satisfied since the cumulants $\kappa$ depend weakly on the mass~\cite{LoVerde:2011iz}.

However, due to the highly nonlinear character of the collapse, one cannot fully parametrise collapsed objects by a single threshold number $\delta_c$. Indeed, it has been shown that the PS prescription does not accurately estimate the halo abundance even in the Gaussian case (for example a small non-spherical perturbation can have a considerable impact)~\cite{Sheth:1999mn}. Hence, the extension to the NG case is guaranteed to also have errors with respect to simulations. What can be done though is to characterise the deviation from Gaussianity by comparing the ratio of G-to-NG densities to that of the PS scheme~\cite{Robinson:1998dx,LoVerde:2007ri}, since the latter is expected to deviate equally in both cases: 
\be \label{mass-function}
\frac{\de n }{\de M} (M,z) = r_{\rm G}(M,z)\frac{\de n_{\rm PS}}{\de M}(M,z),\quad\text{with}\quad r_{\rm G}(M,z)=\dfrac{\frac{\de n }{\de M}\big|_{\rm G}(M,z)}{\frac{\de n_{\rm PS}}{\de M}\big|_{\rm G}(M,z)}.
\ee
For the Gaussian mass function, $\frac{\de n}{\de M}\big|_{\rm G}$, we can adopt a Sheth-Tormen (ST) ansatz~\cite{Sheth:1999mn}, which is better fitted to simulations than the Press-Schechter one, in which case the Gaussian ratio reads
\be \label{gauss-ratio-ST}
r_{\rm G}[\nu_c(z)]=\sqrt{a}A\left(1+(a\nu_c^2)^{-p}\right)e^{\frac{\nu^2_c}2(1-a)},
\ee
with the ST parameters $a = 0.707, \;A = 0.322184,\; p = 0.3$.
With the mass function~\eqref{mass-function} at hand, which is just Eq.~\eqref{mass-function-PS-F} with the replacement $n_{\rm PS}|_{\rm G} \to  n_{\rm ST}$, we may compute the number of clusters per redshift bin above some mass $M$ as~\cite{LoVerde:2007ri}
\be \label{cluster-counts}
\frac{\de N}{\de z}(M,z) = \frac{4\pi}{H(z)}\left(\int\frac{\de z}{H(z)}\right)^2 f_{\rm sky} \int_{M}^\infty \de m \frac{\de n}{\de m}(m,z),
\ee
where $f_{\rm sky}$ is the fraction of the sky covered by each survey. 

This formula can serve as a template for the number density of clusters. For example, in the context of light isocurvature axions one can compute the non-Gaussian PDF deformation $\Delta$~\cite{Chen:2018uul} and thus the mass function $\frac{\de n}{\de M}$ and create mock data via simulations. One can then pick an estimator and devise an overlap between the template and the data, in exactly the same manner that one uses the cosine estimator to measure the overlap of the local template with the actual bispectrum. The simplest thing to do is to take the number counts as an estimator. For example, given a model, one can count how many clusters of mass $M$ exist at redshift $z$ in the mock data and compare this number to the real data.

We may also try to reconstruct the PDF from data using other statistical estimators.
This can be done by simply solving Eq.~\eqref{mass-function}, together with \eqref{F_PS}, as an ODE for the cumulative PDF, $\mu^>$, to get 
\be \label{mu-dn}
\mu^>[\delta_c(z),\sigma_W]= \frac{1 }{2\bar\rho} \int_M^\infty \de m\; \frac{m}{r_{\rm G}[\nu_c(z)]}  \frac{\de n_{\rm }}{\de m}(m,z),
\ee
with $r_{\rm G}$ given by Eq.~\eqref{gauss-ratio-ST}. 
The left hand side now gives the tail distribution of $\delta$, that is, the probability of $\delta>\delta_c$ at redshift $z$. On the right hand side we have the total mass of collapsed objects with $m>M$ at redshift $z$, accounting for corrections to the spherical collapse model via assigning a mass value $m(z)= m_{\rm obs}/r_{\rm G}[\nu_c(z)]$ to an object of observed mass $m_{\rm obs}$ and multiplicity $\frac{\de n_{\rm }}{\de m}(m_{\rm obs},z)$; in principle, this can be deduced from number counts. One can now apply statistical estimators  like Minkowski functionals to the dataset $\left\{\frac{m_{\rm obs}}{r_{\rm G}[\nu_c(z)]},\frac{\de n_{\rm }}{\de m}(m_{\rm obs},z)\right\}$, whose difference from their Gaussian estimate, as in the CMB~\cite{Ade:2013nlj,Buchert:2017uup} and LSS~\cite{Codis:2013exa}, will be a direct probe the NG deformation of the primordial PDF.  Let us note, however, that measuring accurately the mass and redshift of clusters is a hard task, which might complicate a direct connection between cluster counts and primordial NG in this manner.

To summarise, in the context of tomographic NG, one can 1) use Eqs.~\eqref{mass-function-PS-H} and~\eqref{mass-function} as a template for the mass function and thus the number density up to arbitrary order in the Edgeworth expansion; 2) assume a microphysical model, compute the PDF (as in \emph{e.g.} Ref.~\cite{Chen:2018uul}) and use Eqs.~\eqref{mass-function-PS-F} and~\eqref{mass-function} as a template without having to refer to moments; and finally, 3) consider statistical estimators on the LSS dataset probing the primordial PDF via the counting scheme implied by Eq.~\eqref{mu-dn}.

\subsection{Scale dependent halo bias} \label{sec:bias}

Another powerful probe of primordial non-Gaussianity is the halo bias, which enters in the late-time power spectra. As already argued, non-Gaussian initial conditions alter the halo abundance in a nontrivial way by increasing the number of rare density peaks that collapse into halos. This is easier to visualise in the case of the quadratic local ansatz $\delta=\delta_{\rm G}+f_{\rm NL}  \delta_{\rm G}^2$~\cite{Dalal:2007cu}: a positive $f_{\rm NL}$ adds positive skewness to the density distribution; thus, the same probability now corresponds to higher values of $\delta$ with respect to the Gaussian field, leading to more probable enhanced peaks. In~\cite{Dalal:2007cu}, it was shown how this is encoded in a scale dependent correction to the halo bias given by
\be \label{bias-fnl-shift}
\Delta b(k) = 2 \delta_c (b_{\rm G}-1) \frac{f_{\rm NL}}{\alpha(k)},
\ee
where $b_{\rm G}$ is the Eulerian Gaussian bias.
The $k^2$ factor in $\alpha$ (see Eq.~\eqref{poisson}) implies that the effects of non-Gaussianity will be accentuated in large scales (the transfer function goes to 1 for $k\to 0$), which brings surveys like {\sc Ska} and {\sc Lsst} to the frontline of NG searches, promising $\sigma(f_{\rm NL})\lesssim 1$~\cite{Camera:2014bwa,Munchmeyer:2018eey}.
This result was rederived and generalised in~\cite{Matarrese:2008nc,Slosar:2008hx} using different approaches and has been confirmed with N-body simulations (see \emph{e.g.} Refs.~\cite{Baldauf:2015vio,PhysRevD.81.063530,Scoccimarro:2011pz,Desjacques:2008vf,Grossi:2008vf,Pillepich:2008ka}), while in Refs.~\cite{Smith:2011ub,LoVerde:2011iz}, the scale dependent halo bias was computed for the case of cubic $g_{\rm NL},\tau_{\rm NL}$-type local NG ---see Refs.~\cite{Desjacques:2016bnm,Biagetti:2019bnp} for reviews. In what follows we extend it to the generalised local ansatz \eqref{general-ansatz-restricted} and show how in the case of an isocurvature source, future surveys can probe the landscape potential via measurements of the bias factor.

To begin with, let us rewrite the NG ansatz~\eqref{general-ansatz-restricted} for the Newtonian potential $\Phi=\frac35 \zeta$ as 
\be \label{general-loc-Phi}
\Phi(\x) = \phi(\x) + \frac{3}{5} \F_{\rm NG}[\phi](\x),
\ee
where $\phi$ is a Gaussian random field with standard deviation $\sigma_0=\sqrt{\langle \phi^2 \rangle}$. The $3/5$ factor is put explicitly to relate the primordial curvature perturbation $\zeta$ to the late-time (matter-dominated era) gravitational potential $\Phi$.
Following the peak-background split method~\cite{Kaiser:1984sw,Bardeen:1985tr}, we now separate the Gaussian gravitational potential into long and short modes with respect to some characteristic halo scale $R_\star\sim R(M)$ as
\be \label{l/s-phi-split}
\phi = \phi_L + \phi_S,
\ee
which induces a similar split in the variance, $\sigma_0^2=\sigma_L^2+\sigma_S^2 = \langle \phi_L(\x)^2 \rangle + \langle \phi_S(\x)^2 \rangle$. With the help of the expansion in Hermite polynomials~\eqref{F-herm}, and using known identities of these polynomials, we can write
\be \label{F-L-S-expansion-0}
F(\phi_L + \phi_S; \sigma_0^2) = F(\phi_L,\phi_S)  = \sum_{m=0}^\infty \frac{\beta_{m}\left(\phi_L \right)}{m!} \sigma_S^m \,  {\rm He}_m \!\left( \frac{\phi_S}{\sigma_S} \right) ,
\ee
where
\be \label{betas}
\beta_{m}\left(\phi_L \right) \equiv - \sum_{s=0}^\infty \frac{ f_{m+s}}{s!} \sigma_L^s \,  {\rm He}_s \!\left( \frac{\phi_L}{\sigma_L} \right),  
\ee
with $f_0 = f_1 = 0$. How is this expansion of $F$ useful? If one is mainly interested in the short-wavelength dynamics (for instance, to study the gravitational collapse of the matter distribution into galaxies), this expansion allows one to identify the functions $\beta_m$ of long-wavelength fluctuations as effective nonlinearity parameters for the short modes, as may be seen directly from~\eqref{F-L-S-expansion-0}: $m=1$ is a correction to the amplitude of the short modes, $m=2$ a nonlinearity of $f_{\rm NL}$-type, $m=3$ a nonlinearity of $g_{\rm NL}$-type, etc.

We now have to understand quantitatively how long wavelength density fluctuations affect the statistics of short modes and thus the halo number density per unit of halo mass, $\de n/ \de M$, which from now on we denote by $n_L({\x})$, through each term in the expansion~\eqref{F-L-S-expansion-0}. Generically, the halo mass function is a function of the matter contrast and the amplitude of the short modes:
\be
n_L=  n_L[\rho({\x}),\Delta_{\phi_S}].
\ee
Firstly, because of the dependence on $\rho$, irrespective of the presence of non-Gaussianity, a long wavelength perturbation $\delta_L$ will induce a linear background shift in $n$ as 
\be 
n_L[\bar\rho(1+\delta_L({\x}))] = \bar n \left(1 + \frac{\partial \log n_L}{\partial \delta_L} \delta_L({\x}) \right),
\ee
where $\bar n = n_L[\bar\rho]$.
Moving to the non-Gaussian part, the purely long wavelenght contributions ---$m=0$ coefficient in the expansion \eqref{F-L-S-expansion-0}--- by definition will not affect the power spectra, and hence, the bias to first order in the nonlinearity parameters $f_m$, so we may disregard them. From a short-wavelength modes' perspective, they are constant numbers that will only affect the background density through $\delta_L$ in the previous expression. 

Next, we may observe that the coefficient of the term linear in $\phi_S$ ($m=1$) depends on $\phi_L$. Therefore, a long mode will induce a shift in $n({\x})$ through this term, since short modes feel a background perturbed by the local amplitude of the long wavelength perturbation as $\Delta_{\phi_S} \to \left(1+\beta_{1} \! \left(\phi_L \right)\right)\Delta_{\phi_S}$, leading to
\be 
n_L({\x})= \bar n \left(1 + \frac{\partial \log n_L}{\partial \delta_L} \delta_L({\x}) + \beta_{1}\left(\phi_L \right)\frac{\partial \log n_L}{\partial \log\Delta_{\phi_S} }  \right).
\ee
Finally, we need to take into account that the halo density depends on the local amplitude of the long modes through all the functions $\beta_{m}\left(\phi_L \right)$ for all $m \geq 2$, since each $\beta_{m}$ acts as a nonlinearity parameter assuming a local value set by the long wavelength fluctuation $\phi_L$~\cite{Slosar:2008hx,Smith:2011ub}. That is, we should consider the halo density as a function of the form
\be 
n_L({\x}) = n_L[\rho({\x}); \{\beta_m\}_m],
\ee
where $\beta_1$ controls the amplitude of the short modes, and the rest controls the amplitude of their nonlinearities. Thus, we can write
\be \label{n-shift}
n_L({\x}) = 
\bar n \left(1 + \frac{\partial \log n_L}{\partial \delta_L} \delta_L({\x}) + \sum_{m=1} \frac{\partial \log n_L}{\partial \beta_m} \beta_m \! \left(\phi_L (\x) \right)  \right).
\ee

From this expression we can now compute the linear bias, defined as
\be \label{bias-def}
b(k) = \frac{P_{mh}(k)}{P_{mm}(k)},
\ee
where the matter-halo power spectrum is defined as 
\be
P_{mh}(k)=\mathbb{F}_{\x - \y}[ \langle \delta_L(\x) n_L(\y) \rangle] (k),
\ee
and the matter-matter one as 
\be
P_{mm}(k)= \mathbb{F}_{\x - \y}[ \langle \delta_L(\x) \delta_L(\y) \rangle] (k)  =|\delta_k|^2. 
\ee 
Thus, we will need to compute the matter-halo correlator and Fourier transform it ($\mathbb{ F}$). The matter contrast $\delta$ is related ---in subhorizon scales--- to the Newtonian potential through the Poisson equation~\eqref{poisson},
so that $P_{\phi\delta}=P_{\delta\delta}/\alpha$.
The first term in Eq.~\eqref{n-shift} is trivial since it just yields the $\delta$ propagator, which in Fourier space cancels the denominator in Eq.~\eqref{bias-def}, resulting in the standard constant bias, while the second yields the scale dependent correction. Putting everything together, we get
\be \label{linear-b-1}
b(k)  = b_{\rm G} + \frac{1}{\alpha (k)} \sum_{m=1}^{\infty} \frac{\partial \log n_L}{\partial \beta_m} \langle \beta_m'(\phi_L) \rangle = b_{\rm G} - \frac{1}{\alpha (k)} \sum_{m=1}^{\infty} \frac{\partial \log n_L}{\partial \beta_m} f_{m+1} .
\ee 
Evidently, the coefficient of the scale dependent correction contains a summation over all the nonlinearity parameters, or better put, the whole function $F$. In order to see this, we can write it down, equivalently, as
\be \label{linear-b-2}
 b(k) = b_{\rm G} + \frac{1}{\alpha(k)} \int_{-\infty}^{\infty} \mathcal{N}(\phi_L;\sigma_L) \frac{\de F(\phi_L;\sigma_L)}{\de \phi_L} \de \phi_L,
\ee
where $\mathcal{N} \equiv \frac{e^{-\phi_L^2/2\sigma_L^2}}{\sqrt{2\pi} \sigma_L} \sum_{m=1}^\infty \frac{\partial \log n_L}{\partial \beta_m} \frac{1}{\sigma_L^m} {\rm He}_m \! \left( \frac{\phi_L}{\sigma_L} \right) $. Consequently, a scale-dependent halo bias can only signal the presence of \textit{some} form of local NG but not of a specific parametrisation of it.

A downside of this is that the derivatives of $n_L$ w.r.t. $\beta_m$ in Eq.~\eqref{linear-b-2} would have to be computed from simulations if one is to obtain information about $F$, or alternatively, assume a model of collapse into haloes. In the first case, one may compute the derivatives of the halo mass function by varying the initial conditions of the simulations. However, the coefficients~\eqref{betas} of the scale dependent bias cannot be observed in real data in this manner, \emph{i.e}, by repeating the collapse process as one would do in simulations, because they only come once and with the same initial condition for the curvature perturbation field $\zeta$. Therefore, it becomes necessary that the next step be a connection between the effective bias coefficient $\int \mathcal{N} \frac{\de F}{\de \phi} \de\phi $ and observable quantities by modelling the halo mass functional $n_L$ in some way. 

It turns out that the previous section~\ref{sec:halo} provides enough tools to accomplish this. In particular, if one models local NG at short scales $\phi_S$ using the extended PS scheme, as in Eq.~\eqref{mass-function}, the variation with respect to the effective nonlinearity parameter $\de \log n_L/ \de \beta_m$ may be computed directly as $\de \log n_L/ \de f_m^S$, where $f_m^S$ is the local nonlinearity parameter of the short-scale theory (and, to first order in the perturbation $F$, it also defines the corresponding nonlinearity parameter of the full theory because of Eq.~\eqref{F-herm}). This can be done systematically using Eqs.~\eqref{Edgw},~\eqref{mass-function-PS-H} and~\eqref{linear-b-1}.

The first scale dependent NG correction $(m=1)$, associated to cubic NG, was computed in Ref.~\cite{Slosar:2008hx} assuming a universal mass function, in agreement with the result of Ref.~\cite{Dalal:2007cu}:
\be 
\frac{\partial \log n_L}{\partial f_1^S} = 2\delta_c(b_{\rm G}-1), 
\ee
where $b_{\rm G}\equiv\frac{\partial \log n_L}{\partial \delta_L}$ is the Gaussian bias. The second term $(m=2)$, corresponding to a $g_{\rm NL}$-type of NG, was computed in Ref.~\cite{LoVerde:2011iz} using an Edgeworth expansion of the halo mass function and was found to be given by
\be 
\frac{\partial \log n_L}{ \partial \log f_2^S} = \frac{\kappa_3(M)}{6}{\rm He}_3\left[\nu_c(M)\right] - \frac{\kappa'_3(M)}{6\nu_c'(M)}{\rm He}_2\left[\nu_c(M)\right], 
\ee
where $\kappa_n(M)$, is the reduced cumulant defined below Eq.~\eqref{mass-function-PS-H}. It turns out that when~\eqref{linear-b-1} is expanded up to $f_3=g_{\rm NL}$, the resulting expression is in very good agreement with simulations~\cite{LoVerde:2011iz}. Derivatives of the halo mass function with respect to the higher NG nonlinearity parameters $f_4,f_5,\ldots,$ may be easily computed from Eqs.~\eqref{Edgw} and~\eqref{mass-function-PS-H}. Concretely, they are given by
\be
f_m^S \frac{\partial \log n_L}{\partial \beta_m} = \frac{\partial \log n_L}{\partial \log f_m^S} =  \frac{\kappa_{m+1}(M)}{(m+1)!}{\rm He}_{m+1}\left[\nu_c(M)\right] - \frac{\kappa'_{m+1}(M)}{\nu_c'(M) (m+1)!}{\rm He}_{m}\left[\nu_c(M)\right],
\ee
where primes denote derivatives w.r.t. the mass of the halo $M$. These terms can (and should) also be tested with N-body simulations for $f_{m>3}^S$ NG initial conditions, but such a computation is beyond the scope of this paper. 

From Eq.~\eqref{linear-b-1} it is clear that the bias alone cannot differentiate between $f_{\rm NL},\;g_{\rm NL}$ or any higher order NG but it can give an answer to the question of whether Gaussianity is present or not\footnote{A proposal for partially disentangling the contributions of different cumulants is the \emph{density-in-spheres} 1-point PDF~\cite{Uhlemann:2017tex}, which does so by taking into account both overdense and underdense regions, thus breaking the degeneracy between odd/even moments.}. However, given a specific model for primordial NG, this can be a powerful probe in the sense of matching a template against data. For example, in App.~\ref{sec:gla}, we show that one particular realisation of this situation is the presence of an isocurvature mode with a potential $\Delta V(\psi)$. In this case, the function $F$ of the generalised local ansatz~\eqref{general-ansatz-restricted}, is related to the potential as $F \propto \Delta V'$.  
Thus, a measurement of or a constraint on the bias would translate into a constraint on the parameters of the landscape potential. In such a context, one can choose a well-motivated potential $\Delta V$, which fixes the function $F$, depending on few parameters (two or three cover most physically motivated potentials). Then, with the help of simulations, one can use the PDF~\eqref{PDF-windowed} with the replacements~\eqref{z-to-d} to draw appropriate NG initial conditions for the density fluctuation field, and then look for a bias of the form~\eqref{linear-b-1}. For example, within the axion parameter space, there are regions that lead to such cases~\cite{Chen:2018uul} and may be probed in the near future with LSS surveys.

\setcounter{equation}{0}

\section{Concluding remarks} \label{sec:concl}

Non-Gaussianity, notwithstanding how small, is a robust prediction of inflation, and may be detectable if the inflaton had the chance to interact with other degrees of freedom during inflation. Up until now, NG does not show up in low $n$-point correlation functions of the temperature map~\cite{Akrami:2019izv}, at least not in the form of a bispectrum nor a trispectrum, when their estimators are compared to the well-motivated local, equilateral, folded and orthogonal templates. However, primordial NG might be hidden in the data in the form of patterns that need to be revealed through different estimators.

In this work, we have focused on the object that contains the full information about the distribution of anisotropies of the temperature and density fields, that is, the probability density function. Our starting point has been the bottom-up parametrisation of the curvature perturbation as a Gaussian random field plus an arbitrary analytic function thereof. Instead of truncating the series expansion of this function to the first few terms, corresponding to the standard $f_{\rm NL}$, $g_{\rm NL}$ parametrisation, we have kept the entire series enabling us to derive a probability functional that encodes the full local ansatz. 

The various kinds of distributions derived from this functional provide us with new estimators, like the 1-point and 2-point statistics, which constitute complementary channels for the search of NG. In particular, the 2-point PDF encoding the scale dependence of the temperature distribution can serve as a full probe of the power spectrum able to disentangle local NG from other shapes. 
Moreover, the community has been shifting the focus towards LSS statistics as the latter will probe much more modes enhancing the statistical power of the datasets.
With a view towards the near-future LSS surveys, we have computed the distribution of halos of mass $M$ at redshift $z$ showing that it is directly related to the primordial NG deviation. Finally, we computed the NG scale dependent correction to the linear halo bias arguing that it is the full local ansatz that contributes to the scaling. Therefore, the bias offers a unique probe of local NG in full generality, which cannot be used to distinguish the set of nonvanishing nonlinearity parameters, since these contribute as a sum to the correction.

Our results can be used with the future CMB/LSS data in various ways. For example, one can obtain refined constraints on the local ansatz from an Edgeworth expansion of the 2-point CMB temperature PDF. In addition, upon assuming well-motivated microphysical models leading to NG, one can use the formulas for the halo mass function and the bias as templates to be compared with data. Finally, we may use other types of statistical estimators like Minkowski functionals on the CMB/LSS datasets to draw conclusions on the primordial PDF. Let us, nevertheless, stress once more that, as discussed in Sec.~\ref{sec:scnd-order-pt}, our treatment is strictly valid only within linear perturbation theory and as such it is more accurate in cases where primordial NG dominates over secondary effects, especially when one considers higher $n$-point functions, which are well-motivated, for instance, from multi-field effects~\cite{Chen:2018uul, Chen:2018brw} or reheating models~\cite{Bond:2009xx,Suyama:2013dqa}.

Let us finish by pointing out that even though this work is focused on local primordial non-Gaussianity, it should be possible to extend the discussion to other cases. An ``equilateral'' ansatz of the form discussed in the Introduction would lead to a different scale dependence in the 2-point PDF, which could in turn serve as an equilateral NG estimator. Note that constraining such an object via \emph{e.g.} an appropriate Edgeworth expansion could be less costly than that of $n$-point functions using the standard templates. Finally, one may be able to further generalise this by including combinations of spatial derivatives acting on the curvature fluctuation that lead to enfolded correlators. In this more general case, by the same procedure of changing variables in the Gaussian, one would be able to fully describe all the four momentum shapes employed at the level of the PDF.

\section*{Acknowledgements}
We wish to thank Yashar Akrami, Diego Blas, Jonathan Braden, Xingang Chen, Rolando Dunner, Nelson Padilla, Domenico Sapone, Eva Silverstein and Cora Uhlemann for useful discussions and comments. We are also grateful to the organisers of the workshop \emph{Inflation} \& \emph{Geometry} at IAP, Paris, for creating a vibrant atmosphere, which helped us shape part of this work, as well as to all the participants for extensive discussions and feedback. GAP and BSH acknowledge support from the Fondecyt Regular project number 1171811 (CONICYT). BSH is supported by a CONICYT grant number CONICYT-PFCHA/Mag\'{i}sterNacional/2018-22181513. SS is supported by the CUniverse research promotion project (CUAASC) at Chulalongkorn University.

\begin{appendix}

\renewcommand{\theequation}{\Alph{section}.\arabic{equation}}

\setcounter{equation}{0}
\section{Generalised ansatz from quantum fluctuations during inflation} \label{sec:gla}

During inflation, the primordial curvature perturbation $\zeta$ is sourced by quantum fluctuations of the inflaton field or possibly other degrees of freedom such as isocurvaton fields. As a general statement, one can write down the field operator $\zeta(\x)$ at the final time slice $t$ as 
\be
\zeta(\x,t) = U^\dagger(t,t_0) \zeta_I(\x,t) U(t,t_0),
\ee
where $U$ is the temporal evolution operator in the interaction picture of quantum mechanics, and $\zeta_I$ the interaction-picture field, which follows the dynamics of the free theory. 

Naturally, the field $\zeta$ will generate a specific set of $n$-point functions when computing expectation values.  
One can then construct a PDF $\PDF[\zeta]$ that generates these statistics through functional integration
\be
\langle \zeta(\x_1) \cdots \zeta(\x_n) \rangle = \int D \zeta \PDF[\zeta] \zeta(\x_1) \cdots \zeta(\x_n),
\ee
over the field configurations $\zeta(\x)$.  Quantum mechanics does provide the tools to determine $\PDF[\zeta]$ directly, at least in principle. The operation $\zeta = U^\dagger \zeta_I U$ can be reframed in terms of a functional expression
\be
\zeta(\x) = \mathcal{O}[\zeta_I, \Pi_I, \{\psi^I_i, \Pi^I_i\}_i ](\x),
\ee
that depends on the whole spacetime evolution of the interaction-picture fields, $\zeta_I$ and other degrees of freedom $\psi_i^I$, and that of their conjugate momenta, $\Pi_I$ and $\Pi_i^I$. In principle, one can compute correlations directly from this expression. However, if the dependence of $\mathcal{O}$ on the interaction-picture fields is known, one can determine the PDF of $\zeta$ by integrating over all possible configurations:
\be
\PDF[\zeta] \!= \!\!\!\int \!\!D\zeta_I D\Pi_I \left( \prod_i D \psi_i^I D \Pi_i^I \right) \! {\rm P}_{\rm G}[\{\psi^I_i, \Pi^I_i\}_i] \left( \prod_\x \delta \left( \zeta(\x) - \mathcal{O}[\zeta_I, \Pi_I, \{\psi^I_i, \Pi^I_i\}_i ](\x) \right) \right),
\ee
where ${\rm P}_{\rm G}$ is a Gaussian measure, with appropriate prescriptions to take into account the ordering of operators in $\mathcal{O}$. The measure is guaranteed to be Gaussian because the free fields evolve linearly in time, and therefore the contraction of quantum field operators obeys Wick's theorem, which is equivalent to saying that the statistics are Gaussian. 

Once $\PDF[\zeta]$ is obtained, the statistics of $\zeta$ is fully determined. Computationally, however, it is useful to have a probability distribution from which one knows how to obtain expectation values. On the other hand, one knows that the observed statistics for $\zeta$ are consistent with Gaussianity, and that deviations, if any, must be small. This motivates finding a functional map $\zeta_{\rm G}(\x) = G[\zeta](\x)$, with inverse $\zeta(\x) = \zeta_{\rm G}(\x) + \mathcal{F}[\zeta_{\rm G}](\x)$, such that $\zeta_{\rm G}$ has Gaussian statistics, \emph{i.e.}, such that
\be
\PDF_{\rm G}[\zeta_{\rm G}] = \PDF[\zeta_{\rm G} + \F_{\rm NG}[\zeta_{\rm G}]]  \times {\rm det} \left( \frac{\delta \zeta}{\delta \zeta_{\rm G}} \right).
\ee
The difficulty, of course, lies in finding such a mapping. Afterward, one can include another mapping, that makes the power spectrum of $\zeta_{\rm G}$ to be consistent with current observations, if this is not the case already.

Thus, we have justified writing $\zeta$ in terms of a Gaussian field $\zeta_{\rm G}$,
\be
\zeta(\x) = \zeta_{\rm G}(\x) + \mathcal{F}[\zeta_{\rm G}](\x),
\ee
where the functional $\F_{\rm NG}$ is, in principle, arbitrary, and should be determined from the specifics of the inflationary model at hand. We now give an example of a model that motivates searching for non-Gaussianity wherein
\be
\F_{\rm NG}[\zeta_{\rm G}](\x) = \int_\y \int_\K e^{i \K \cdot (\x - \y)} F(\zeta(\y)),
\ee
which is the main focus of our work.

\subsection{A concrete example: multi-field inflation} \label{sec:axion}

Our current understanding of fundamental theories, such as string theory and supergravity, requires us to take into consideration the existence of many scalar fields. In such a framework, it does not make much sense to talk about an inflaton field; instead, one has an inflationary path meandering through a landscape potential. Consequently, curvature fluctuations may have been coupled to (many) other dynamically active degrees of freedom inevitably sourcing departures from Gaussianity. Even in the simplest extension ---from single-field to multi-field inflation--- one finds a rich phenomenology that can be tested by near-future surveys. The presence of multiple fields during inflation leaves unique imprints in the CMB, a subject that has gained a lot of attention in the last years with quasi-single field inflation~\cite{Chen:2009we, Chen:2009zp} and the Cosmological Collider program~\cite{Arkani-Hamed:2015bza}. The ``smoking gun'' signature of massive degrees of freedom active at the inflationary energy scale is a bispectrum of the local shape that peaks for triplets of modes with one momentum going to zero.

In the case of two-field models of inflation, the Lagrangian describing the dynamics of the curvature fluctuation $\zeta$ interacting with an isocurvature field $\psi$ is found to have the following generic form:
\be
\mathcal L = a^3 \Big[\epsilon  ( \dot \zeta - \alpha \psi ) ^2 -   \frac{\epsilon}{a^2} (\nabla \zeta)^2  +  \frac{1}{2} \dot \psi^2  - \frac{1}{2a^2} (\nabla \psi)^2    - \frac{1}{2} \mu^2 \psi^2 \Big] + \mathcal L_{\rm int}  . \label{Lagrantian-full}
\ee
In the previous expression, $a$ is the scale factor, $\epsilon = - \dot H / H^2$ is the first slow-roll parameter (where $H = \dot a / a$ is the Hubble expansion rate during inflation), $\mu$ is the so-called entropy mass of the isocurvature field $\psi$, and $\alpha$ is a coupling that appears as a consequence of the shape of the background inflationary path in the multi-field space. It parametrises the bending of inflationary paths in field space~\cite{Gordon:2000hv, GrootNibbelink:2000vx, GrootNibbelink:2001qt}, with $\alpha = 0$ corresponding to the case of geodesic trajectories (a straight line in field space). The piece $\mathcal L_{\rm int}$ contains operators of cubic order (and higher) in terms of the fields $\zeta$ and $\psi$.  A nonvanishing value of $\alpha$ makes $\psi$ act as a source for the amplitude of $\zeta$ via the quadratic mixing term
\be
\mathcal L_{\rm mix}^{(2)} = - 2 \epsilon a^3 \alpha \dot \zeta \psi .  \label{mix-quadratic}
\ee
Crucially, this mixing allows the transfer of non-Gaussian statistics from the isocurvature field to the curvature fluctuation (but not the other way around~\cite{Achucarro:2016fby}). The generated non-Gaussianity will depend on the particular form of $\mathcal L_{\rm int}$. 

A standard approach to deal with~(\ref{Lagrantian-full}) consists in assuming that $\mathcal L_{\rm int}$ can be  organised perturbatively in terms of powers of both fields $\zeta$ and $\psi$. Such an approach supposes that the nonlinear dynamics is dominated by cubic operators $\mathcal L_{\rm int}  \supset  \mathcal L_{\rm int}^{(3)}$, and implies that the main non-Gaussian departures will be parametrised by the bispectrum, the amplitude of the 3-point function in momentum space~\cite{Gangui:1993tt, Komatsu:2001rj, Acquaviva:2002ud, Maldacena:2002vr, Bartolo:2004if, Liguori:2010hx, Chen:2010xka, Wang:2013eqj}. Within this approach, one finds that the main operators leading to a nonvanishing bispectrum are given by a self-interaction term and a mixing term of the forms $\mathcal L_{\rm self}^{(3)} \propto a^3 g \psi^3$ and $ \mathcal L_{\rm mix}^{(3)} \propto a^3 \alpha \dot \zeta \psi^2$, respectively.  Any other term in $\mathcal L_{\rm int}^{(3)}$ turns out to be suppressed by powers of the slow-roll parameters. Both interaction terms lead to a non-analytic behaviour of the bispectrum: in the $\mu/H<3/2$ regime, one obtains an NG amplitude enhanced by $g$, with the bispectrum following an irrational power law in the squeezed limit leading to an intermediate shape between the equilateral and local ones~\cite{Chen:2009we, Chen:2009zp}; the other region $\mu/H>3/2$ leads to a bispectrum with a shape sensitive to the value of the entropy mass, displaying oscillatory patterns~\cite{Noumi:2012vr, Chen:2015lza, Chen:2016cbe, Lee:2016vti}, while in the $\mu / H \gg 3/2$ limit the heavy field may be fully integrated out, leading to equilateral shape NG~\cite{Gong:2013sma}. 

However, one may conceive regimes characterised by interaction Lagrangians $\mathcal L_{\rm int}$ in which one is not allowed to disregard terms of higher powers in the fields. For instance, the field $\psi$ may have a potential $\Delta V(\psi)$ displaying a rich structure within a wide field range $\Delta \psi$~\cite{Palma:2017lww}. In this case, the Lagrangian (\ref{Lagrantian-full}) can be rewritten as
\be
\mathcal L = a^3 \Big[\epsilon  ( \dot \zeta - \alpha \psi ) ^2 -   \frac{\epsilon}{a^2} (\nabla \zeta)^2  +  \frac{1}{2} \dot \psi^2  - \frac{1}{2a^2} (\nabla \psi)^2    - \Delta V (\psi) \Big]  + \cdots , \label{Lagrantian-full-2}
\ee
where we have absorbed the mass term in $\Delta V(\psi)$ by means of the identification $\mu^2 \equiv \Delta V'' (0)$, where primes denote derivatives with respect to $\psi$. The elipses $\cdots$ denote terms that are either suppressed by slow-roll parameters or by $\alpha/H$, which is assumed to be small. This Lagrangian (\ref{Lagrantian-full-2}) describes the dynamics of perturbations in those cases where the landscape potential has a rich structure in directions orthogonal to the inflationary path. If the potential remains shallow, \emph{i.e.} $\Delta V / H^4 \ll 1$, the field $\psi$ will display characteristic fluctuations $\Delta \psi$ that could traverse many minima and maxima of the potential, therefore probing the structure of the landscape. In this case, one cannot just stick to computations involving the first two terms of the expansion $\Delta V = \frac{\mu^2}{3} \psi^2 + \frac{g}{3!} \psi^3 + \cdots$; instead, one has to perform perturbative computations where information about the entire potential $\Delta V$ is kept under control. As shown in~\cite{Chen:2018uul, Chen:2018brw}, as long as $\Delta V / H^4 \ll 1$, it is indeed possible to compute every $n$-point correlation function for $\zeta$ and derive its probability distribution function, $\PDF (\zeta)$. Remarkably, this PDF  turns out to be given by a Gaussian profile with small non-Gaussian corrections determined by the shape of the landscape potential $\Delta V$. Thanks to this property, it is in principle possible to reconstruct the shape of the landscape potential $\Delta V$ out of CMB observations~\cite{Chen:2018brw}, providing information about a section of $\Delta V$ during the period of inflation when the modes of wavelengths relevant to the CMB were exiting the horizon (hence the term, tomographic non-Gaussianity) and beyond.

The generation of tomographic non-Gaussianity may be traced back to the self-interactions of an isocurvature field $\psi$. These self-interactions are transferred to the curvature perturbations thanks to a linear interaction coupling both fields [see Eq.~(\ref{Lagrantian-full-2})]. To appreciate how self-interactions give rise to this form of non-Gaussianity, we may start by recalling that in the interaction picture, the evolution of the field $\psi (\x,\tau)$ is given by
\be
\psi (\x,\tau) = U (\tau,\tau_0) \psi_I (\x,\tau) U^\dag (\tau,\tau_0).
\ee 
From a quantum-mechanical perspective in the Heisenberg picture, one can write
\be
\psi(\x,\tau) = \psi(\x, \tau_0) + i \int^{\tau}_{\tau_0}  \!\!\! \de \tau' [H(\tau'), \psi(\x,\tau')].
\ee
Similarly, the interaction picture gives that to first order in the interaction the field is given by
\be
\psi(\x,\tau) = \psi_I(\x, \tau) + i \int^{\tau}  \!\!\!  \de\tau' [H_I(\tau'), \psi_I(\x,\tau)],
\ee
where the subscript $I$ informs us that the corresponding operator is in the interaction picture and evolves as a free field. If, for simplicity, we take de Sitter spacetime as a background, the interaction-picture Hamiltonian reads $H_I(\tau) = \int_x a^4(\tau) \Delta V(\psi_I(\x,\tau))$ with $a(\tau) = -1/H\tau$. With this, we may work on our previous equation to obtain
\be
\psi(\x,\tau) = \psi_I(\x, \tau) + i \int^{\tau}  \!\!\!  \de\tau' \! \int_x  a^4(\tau') \, [\psi_I(\x',\tau'),\psi_I(\x,\tau)] \frac{\partial \Delta V}{\partial \psi}(\psi_I(\x',\tau')).
\ee
With the help of canonical commutation relations for the appropriate field $v \equiv a \psi$~\cite{Achucarro:2016fby,Chen:2018uul}, the commutator $[\psi_I(\x',\tau'),\psi_I(\x,\tau)]$ is just a number and we can carry out the integral over $\tau'$ explicitly provided that:
\begin{enumerate}
\item The quantum field $\psi_I(\x',\tau')$ in the argument of $\partial \Delta V/\partial \psi$ may be treated as a constant over time $\tau'$.
\item The range of modes under consideration involves superhorizon modes with $|k\tau'| \lesssim 1$.
\end{enumerate}
In fact, if the range of modes satisfies $|k\tau'| \ll 1$ for all $k$, then the first condition is implied by the second. 

Let us give some comments about these conditions: The first point seems natural in the sense that the statistics of $\psi_I$ do not evolve over time: it may be seen as a Gaussian random field with a definite covariance. The second point, although appealing, is both physically and mathematically suspect, since in principle the interaction-picture Hamiltonian involves every mode (\emph{i.e.} every momentum scale). Nonetheless, from an EFT perspective this is perfectly acceptable, as long as the potential $\Delta V$ is responsible for describing the physics at those scales. Moreover, this is the appropriate course of action when studying CMB or LSS modes that spent a large number of e-folds outside the horizon, because they do satisfy $|k\tau'| \ll 1$ throughout most of their history (practically for every time after horizon crossing this condition is fulfilled).

Using these considerations, one obtains
\be \label{local-isocurvature}
\psi(\x,\tau) = \psi_I(\x, \tau) - \frac{\Delta N}{3H^2} \int_\y \int_{\K} e^{i\K \cdot (\x - \y) } \frac{\partial \Delta V}{\partial \psi}(\psi_I(\y,\tau)),
\ee
where $\Delta N$ is the number of e-folds spent outside the horizon by the range of modes under consideration, which we take to satisfy $\Delta N \gg 1$. While the former may seem to be a heavy restriction, if we consider that the currently observable range of scales in the CMB satisfies $\ln (k_{S}/k_L) \simeq 8$~\cite{Akrami:2018odb} and $\Delta N \sim 60$, we see that approximating $\Delta N$ by a single value for all modes in the considered range is justified. Of course, as was earlier suggested, a rigorous proof of this requires to go through every $n$-point function, performing an adequate renormalisation procedure to make the computations consistent at every bounded range of momenta. This was thoroughly dealt with in previous works~\cite{Chen:2018uul, Chen:2018brw}, although stopping just short of writing down Eq.~\eqref{local-isocurvature}. 

The curvature perturbation $\zeta(\x,\tau)$ may be obtained in a completely analogous manner, only that we now have to consider an extra commutator to account for the quadratic mixing term~\eqref{mix-quadratic}:
\be \label{zeta-perturbated}
\zeta (\x,\tau) = \zeta_I(\x,\tau) + i\int^{\tau} \!\!\! \de\tau'  \, [H_I^{\alpha}(\tau'), \zeta_I(\x,\tau)]  - \int^{\tau} \!\!\! \de\tau' \!  \int^{\tau'} \!\!\!\!\! \de\tau'' \, [H_I^{V}(\tau''), [H_I^{\alpha}(\tau'), \zeta_I(\x,\tau)]] + \cdots,
\ee
where $H^{V}_I$ is the term of the interaction-picture Hamiltonian that contains the $\psi$ self-interactions and $H^{\alpha}_I$ contains terms associated to the quadratic mixing. The ellipsis $\cdots$ stand for higher order terms. It is of crucial importance to obtain the correct result to notice that the commutators in the last term of~\eqref{zeta-perturbated} only give a nonzero result when the pieces of the interaction Hamiltonian are written in that order. This, alongside the time ordering, yields an additional $1/2$ factor for the statistical transfer of the nonlinear perturbation $\Delta V$. After a calculation analogous to the one that led us to~\eqref{local-isocurvature}, with the same working assumptions, one obtains
\be \label{zeta-perturbated-2}
\zeta (\x,\tau) = \zeta_I(\x,\tau) + \frac{\alpha}{H} \Delta N \left( \psi_I (\x,\tau) - \frac{1}{2} \frac{\Delta N}{3 H^2} \int_\y \int_{\K} e^{i\K \cdot (\x - \y) }  \Delta V' \left( \psi_I (\y,\tau) \right) \right).
\ee
This equation can accommodate a variety of regimes. However, we choose to work in a situation wherein  the linear transfer from $\psi$ to $\zeta$ dominates. Even though this last equation is a perturbative result, $\zeta = \frac{\alpha \Delta N}{H} \psi + \zeta_0 $ is an exact solution of the equations of motion on superhorizon scales~\cite{Achucarro:2016fby}. This allows us to neglect the first term. Furthermore, since conventionally $\zeta$ should satisfy $\langle \zeta \rangle = 0$, from this point forward we will consider
\be \label{zeta-perturbated-3}
\zeta (\x,\tau) \approx \frac{\alpha}{H} \Delta N \left( \psi_I (\x,\tau) - \frac{1}{2} \frac{\Delta N}{3 H^2} \int_\y \int_{\K} e^{i\K \cdot (\x - \y) }  [\Delta V' \left( \psi_I (\y,\tau) \right) - \langle \Delta V' \left( \psi_I (\y,\tau) \right) \rangle ] \right),
\ee
or equivalently, omitting the temporal coordinate and the projection integration $\int_\y \int_{\K} e^{i\K \cdot (\x - \y) }$,
\be \label{zeta-perturbated-4}
\zeta (\x) = \zeta_{\rm G} (\x) - \frac{\alpha^2 \Delta N^2}{2H^2} \frac{\Delta N}{3 H^2}  \left[ \frac{\partial}{\partial \zeta} \left( \Delta V \left( \frac{H \zeta_{\rm G} (\x)}{\alpha \Delta N} \right) \right) - \left\langle \frac{\partial}{\partial \zeta} \left( \Delta V \left( \frac{H \zeta_{\rm G} (\x)}{\alpha \Delta N} \right) \right) \right\rangle \right],
\ee
where we have written $\zeta_{\rm G}$ instead of $\psi_I$ to stress the nature of our result: $\zeta$ is made up from a Gaussian contribution plus a local non-Gaussian term. This result has the desired form $\zeta(\x) = \zeta_{\rm G}(\x) + \F_{\rm NG} [\zeta_{\rm G}] (\x)$. To identify $\F_{\rm NG}$ it is convenient to recognise that, because at linear order $\zeta \simeq \frac{\alpha \Delta N}{H} \psi$, the power spectrum of $\zeta$ satisfies $P_\zeta = \frac{\alpha^2 \Delta N^2}{H^2} P_{\psi} $, where $k^3 P_{\psi} / 2 \pi^2= H^2 / 4 \pi^2$ (because in the free theory $\psi$ behaves as a massless field). This allows one to find~\cite{Chen:2018brw} $\alpha^2 \Delta N^2 = 4 \pi^2 A_s$ where $A_s =  k^3 P_\zeta / 2 \pi^2$ is the amplitude of the power spectrum $P_\zeta$ of $\zeta$. This finally leads to
\be \label{F-V-app}
\F_{\rm NG} (\zeta) \propto \frac{\partial }{\partial \zeta} \Delta V \left( \frac{H}{2\pi A_s^{1/2}} \zeta \right) ,
\ee
which is the result\footnote{A similar expression with a generic NG deviation parametrised by some function was written in Refs.~\cite{Bond:2009xx,Suyama:2013dqa} in the context of preheating. Here we see that another natural interpretation of such non-Gaussianity is an isocurvature potential, with the trigonometric case corresponding to an axion~\cite{Chen:2018uul}.} reported in the Introduction.

\setcounter{equation}{0}
\section{The partition function} \label{sec:Partition}

The defining property of a partition function is that upon functional differentiation as in Eq.~\eqref{Z-corr}, it should give the $n$-point functions:
\be
\langle \zeta({\K}_1 ) \cdots \zeta({\K}_n )\rangle = \frac{\delta^n Z[J]}{(i\delta J(-{\K}_1 )) \cdots (i \delta J(-{\K}_n)) }\bigg|_{J = 0} = \int D\zeta \PDF[\zeta] \zeta({\K}_1 ) \cdots \zeta({\K}_n ).
\ee
This is accomplished by the following functional:
\be \label{Z-Fourier-app}
Z[J] = \int D \zeta \, \PDF[\zeta] \, e^{i\int_\K \zeta({\K}) J(-{\K})} = \int D \zeta \, \PDF[\zeta] \, e^{i\int_\y \zeta({\y}) J({\y})}.
\ee
To evaluate this expression, let us use the expression for $\PDF[\zeta]$ as given in Eq.~(\ref{PDF}):
\begin{align}
\PDF_F[\zeta] =& \, \PDF_{\rm G}[\zeta] \times \left[ 1 - \int_\x  \int_{\K} \frac{\partial F}{\partial \zeta} (\zeta(\x)) + \int_\x \int_\y \zeta(\x) \int_{\K} \frac{e^{i \K \cdot (\x - \y)} }{P_\zeta(k)}  \int_\z \int_{\bf q} e^{i {\bf q} \cdot (\y - \z)} F(\zeta(\z))   \right] \nn  \\
=& \,  \PDF_{\rm G}[\zeta] \times \left[ 1 - \int_\x  \int_{\K} \frac{\partial F}{\partial \zeta} (\zeta(\x)) + \int_\x  \int_{\K} \int_\z \zeta(\x)  \frac{e^{i \K \cdot (\x - \z)} }{P_\zeta(k)} F(\zeta(\z))   \right].
\end{align}
We will now evaluate~\eqref{Z-Fourier-app} by expanding $F$ in a power series and using Wick's theorem. 
Since~\eqref{Z-Fourier-app} may be read as computing the expectation value of
\begin{align} \label{expectand}
 e^{i\int_\y \zeta({\y}) J({\y})} \left[ 1 - \int_\x  \int_{\K} \frac{\partial F}{\partial \zeta} (\zeta(\x)) + \int_\x  \int_{\K} \int_\z \zeta(\x)  \frac{e^{i \K \cdot (\x - \z)} }{P_\zeta(k)} F(\zeta(\z))   \right]
\end{align}
over a Gaussian measure, we will do exactly that. Therefore, we will have to compute
\be \label{expectands}
\left\langle e^{i\int_\y \zeta({\y}) J(\y)} \frac{\partial F}{\partial \zeta} (\zeta(\x))  \right\rangle_{\rm G} \qquad {\rm and} \qquad \left\langle e^{i\int_\y \zeta({\y}) J(\y)}  \zeta(\x) F(\zeta(\z))  \right\rangle_{\rm G},
\ee
where the subscript $G$ instructs to take the expectation value over a Gaussian measure. Note that the second expectation value, per Wick's theorem, can be written as the sum of two expectation values
\be
\begin{split}
\left\langle e^{i\int_\y \zeta({\y}) J({\y})}  \zeta(\x) F(\zeta(\z))  \right\rangle_{\rm G} =& \, i \int_\w J(\w) \left\langle \zeta(\w) \zeta(\x)  \right\rangle_{\rm G} \left\langle e^{i\int_\y \zeta({\y}) J({\y})} F(\zeta(\z))  \right\rangle_{\rm G} \\ 
& + \left\langle  \zeta(\x) \zeta(\z)  \right\rangle_{\rm G} \left\langle e^{i\int_\y \zeta({\y}) J({\y})}  \frac{\partial F}{\partial \zeta}(\zeta(\z))  \right\rangle_{\rm G} \\
=& \, i \int_\w J(\w) \Sigma(\w,\x) \left\langle e^{i\int_\y \zeta({\y}) J({\y})} F(\zeta(\z))  \right\rangle_{\rm G} \\ 
& + \Sigma(\x,\z) \left\langle e^{i\int_\y \zeta({\y}) J({\y})}  \frac{\partial F}{\partial \zeta}(\zeta(\z))  \right\rangle_{\rm G}.
\end{split}
\ee
Now, using that $\Sigma(\x,\z) = \int_\q P_\zeta(q) e^{i \q \cdot(\x - \z) }$, and replacing the last term into the corresponding term of~\eqref{expectand}, we get
\be
\int_\x  \int_{\K} \int_\z \frac{e^{i \K \cdot (\x - \z)} }{P_\zeta(k)}  \int_\q P_\zeta(q) e^{i \q \cdot(\x - \z) } \left\langle e^{i\int_\y \zeta({\y}) J({\y})}  \frac{\partial F}{\partial \zeta}(\zeta(\z))  \right\rangle_{\rm G}.
\ee 
Then, integrating over $\x$ gives $|\q| = |\K|$, and thus yields
\be
\int_\z  \int_{\K} \left\langle e^{i\int_\y \zeta({\y}) J({\y})}  \frac{\partial F}{\partial \zeta}(\zeta(\z))  \right\rangle_{\rm G},
\ee
which is equal but opposite in sign to the first term of~\eqref{expectands}. Therefore, those two cancel out, and we only have to compute
\be
\int_\x  \int_{\K} \int_\z \frac{e^{i \K \cdot (\x - \z)} }{P_\zeta(k)} i \int_\w J(\w) \Sigma(\w,\x) \left\langle e^{i\int_\y \zeta({\y}) J({\y})} F(\zeta(\z))  \right\rangle_{\rm G}.
\ee
If we define
\be
F(\zeta) = \sum_{n=0}^\infty \frac{g_n}{n!} \zeta^n,
\ee
then the object of interest in the computation is
\be
\frac{i^m g_n}{m! n!} \left\langle \left( \int_\y \zeta({\y}) J({\y})\right)^m \zeta(\z)^n  \right\rangle_{\rm G}.
\ee
There are three type of contractions in this expression: two self-contractions of fields originating from equivalent expressions and a mixed one. Performing the combinatorics gives
\be
\begin{split}
\frac{i^m g_n}{m! n!} \sum_{\substack{m',n',\ell' \\ 2m' + \ell' = m \\ 2n' + \ell' = n }} & \frac{m!}{2^{m'} m'! (m - 2m' - \ell')! } \left( \int_x \int_y J(\x) \Sigma(\x,\y) J(\y) \right)^{m'} \\ 
& \times \frac{1}{\ell'!} \left( \int_\y J(\y) \Sigma(\y,\z) \right)^{\ell'}  \times \frac{n!}{2^{n'} n'! (n - 2n' - \ell')! } \left( \Sigma(\z,\z) \right)^{n'},
\end{split}
\ee
which one can sum over $n,m$ to eliminate the constraints on the $m',n',\ell'$ sums. This gives
\begin{align}
& \left\langle e^{i\int_\y \zeta({\y}) J({\y})} F(\zeta(\z))  \right\rangle_{\rm G} \nn \\
&= \sum_{n',m',\ell'=0}^\infty \frac{i^{2m' + \ell'} g_{2n' + \ell}}{2^{m'} m'! 2^{n'} n'! \ell'!} \left( \int_x \int_y J(\x) \Sigma(\x,\y) J(\y) \right)^{m'}  \left( \int_\y J(\y) \Sigma(\y,\z) \right)^{\ell'}  \left( \Sigma(\z,\z) \right)^{n'} \nn \\
&= \exp \left[- \frac{1}{2} \int_x \int_y J(\x) \Sigma(\x,\y) J(\y) \right] \sum_{n',\ell'=0}^\infty g_{2n'+\ell'} \frac{1}{n'!} \left( \frac{1}{2} \Sigma(\z,\z) \right)^{n'} \frac{1}{\ell'!} \left(i \int_\y J(\y) \Sigma(\y,\z) \right)^{\ell'} \nn \\
&=  e^{- \frac{1}{2} \int_x \int_y J(\x) \Sigma(\x,\y) J(\y)} \sum_{n',\ell'=0}^\infty \frac{\left(i \int_\y J(\y) \Sigma(\y,\z) \frac{\partial }{\partial \zeta}  \right)^{\ell'}}{\ell'!}   \frac{\left( \frac{1}{2} \Sigma(\z,\z) \frac{\partial^2 }{\partial \zeta^2} \right)^{n'}}{n'!}  \left( \sum_{n=0}^\infty \frac{ g_{n}}{n!} \zeta^n \right) \bigg|_{\zeta = 0} \nn \\
&=  e^{- \frac{1}{2} \int_x \int_y J(\x) \Sigma(\x,\y) J(\y)}   \exp \left[\frac{\Sigma(\z,\z)}{2} \frac{\partial^2}{\partial \zeta^2} \right]   F(\zeta)  \bigg|_{\zeta = i \int_\y J(\y) \Sigma(\y,\z)}.
\end{align}
Identifying $\sigma_\zeta^2 = \Sigma(\z,\z)$ and using the Weierstrass transform, we get
\begin{align}
& \int_\x  \int_{\K} \int_\z \frac{e^{i \K \cdot (\x - \z)} }{P_\zeta(k)} i \int_\w J(\w) \Sigma(\w,\x) \left\langle e^{i\int_\y \zeta({\y}) J({\y})} F(\zeta(\z))  \right\rangle_{\rm G} \nn \\
&= i e^{- \frac{1}{2} \int_x \int_y J(\x) \Sigma(\x,\y) J(\y)} \int_\z J(\z) \!\!  \int_{-\infty}^\infty \!\!\!\! \de\zeta \frac{\exp \left\{- \frac{ \left( \zeta - i \int_\y J(\y) \Sigma(\y,\z) \right)^2 }{2 \sigma_\zeta^2} \right\} }{\sqrt{2\pi \sigma_\zeta^2}} F(\zeta) .
\end{align}
Finally, we use that
\begin{align}
 \frac{\partial }{\partial \zeta} e^{- \frac{ \left( \zeta - i \int_\y J(\y) \Sigma(\y,\z) \right)^2 }{2 \sigma_\zeta^2} } &= e^{- \frac{ \left( \zeta - i \int_\y J(\y) \Sigma(\y,\z) \right)^2 }{2 \sigma_\zeta^2} } \left( i \int_\y J(\y) \Sigma(\y,\z) - \zeta \right) \frac{1}{\sigma_\zeta^2} \nn \\
 \implies e^{- \frac{ \left( \zeta - i \int_\y J(\y) \Sigma(\y,\z) \right)^2 }{2 \sigma_\zeta^2} } &= \frac{1}{i \int_\y J(\y) \Sigma(\y,\z)} \left( \zeta + \sigma_\zeta^2 \frac{\partial }{\partial \zeta}  \right) e^{- \frac{ \left( \zeta - i \int_\y J(\y) \Sigma(\y,\z) \right)^2 }{2 \sigma_\zeta^2} },
\end{align}
to get, after integration by parts,
\begin{align}
& \int_\x  \int_{\K} \int_\z \frac{e^{i \K \cdot (\x - \z)} }{P_\zeta(k)} i \int_\w J(\w) \Sigma(\w,\x) \left\langle e^{i\int_\y \zeta({\y}) J({\y})} F(\zeta(\z))  \right\rangle_{\rm G} \nn \\
&=  e^{- \frac{1}{2} \int_x \int_y J(\x) \Sigma(\x,\y) J(\y)} \!\! \int_\z \frac{J(\z)}{\int_\y J(\y) \Sigma(\y,\z)} \!\!  \int_{-\infty}^\infty \!\!\!\! \de\zeta \frac{\exp \left\{- \frac{ \left( \zeta - i \int_\y J(\y) \Sigma(\y,\z) \right)^2 }{2 \sigma_\zeta^2} \right\} }{\sqrt{2\pi \sigma_\zeta^2}} \left( \zeta - \sigma_\zeta^2 \frac{\partial^2}{\partial \zeta^2} \right) F(\zeta) \nn \\
&=  \exp \left[- \frac{1}{2} \int_k \int_y |J(\K)|^2 P_\zeta(k) \right] \nn \\ 
& \,\,\,\, \times \int_\x \frac{\int_\K e^{i\K \cdot \x} J(-\K)}{\int_\K  e^{i\K \cdot \x} J(-\K)  P_\zeta(k) } \!\!  \int_{-\infty}^\infty \!\!\!\! \de\zeta \frac{\exp \left\{- \frac{ \left( \zeta - i \int_k e^{i\K \cdot \x} J(-\K) P_\zeta(k) \right)^2 }{2 \sigma_\zeta^2} \right\} }{\sqrt{2\pi \sigma_\zeta^2}} \left( \zeta - \sigma_\zeta^2 \frac{\partial^2}{\partial \zeta^2} \right) F(\zeta),
\end{align}
which gives
\be 
\begin{split}
Z[&J]  = \, \exp \left[-\frac{1}{2} \int_\K J(\K) J(-\K) P_\zeta(k) \right] \;\times\\  &  \left( 1 - \int_\x \frac{\int_\K e^{i\K \cdot \x} J(-\K)}{ \int_\K e^{i\K \cdot \x} J(-\K) P_\zeta(k) } \int_{ \zeta} \frac{\exp \left[ - \frac{\left({ \zeta} - i\int_\K e^{i\K \cdot \x} J(-\K) P_\zeta(k) \right)^2}{2 \sigma_\zeta^2} \right] }{\sqrt{2 \pi} \sigma_\zeta} \left(\sigma_\zeta^2 \frac{\partial }{\partial \zeta} - {\zeta}\right)  F( \zeta) \right),
\end{split}
\ee
as shown in the main text.

\setcounter{equation}{0}
\section{Details of the fixed-point PDFs} \label{sec:2-point-det}

\subsection{2-Point PDF: general case}

In the main text we argued that the 2-point PDF can be deduced from the probability density functional upon conditioning in two points. Here we arrive at the same result starting from the corresponding 2-point correlators and deriving the PDF. To our knowledge, this has not been derived earlier and thus we outline the procedure in some detail. 

As in the previous case, the PDF must include the non-fully connected contributions. This is a combinatorial mess, for we expect
\be
\begin{split}
\langle \zeta^{n_1}(\x_1) \zeta^{n_2}(\x_2) \rangle = \sum_{m_1, m_2, m_t} & \#_{n_i,n_2,m_1,m_2, m_t} \left(\sigma_{\zeta}^2|_{\x_1}\right)^{m_1} \left(\sigma_{\zeta}^2(x) \right)^{m_t} \left(\sigma_{\zeta}^2|_{\x_2}\right)^{m_2} \\ & \times \langle \zeta^{n_1 - 2m_1 - m_t}(\x_1) \zeta^{n_2 - 2m_2 - m_t}(\x_2) \rangle_c.
\end{split}
\ee
Let us calculate $\#_{n_1,n_2,m_1,m_2,m_t}$: we have an overall factor $n_1! n_2!$ from which we must divide out the overcounted terms. In this counting, we have $m_t!$ redundant permutations when connecting $\x_1$ and $\x_2$, plus $2^{m_1} m_1! 2^{m_2} m_2!$ when pairing amongst themselves. Finally, the ones that are assigned to the fully connected contribution undergo no further permutation, thus we must also divide by $(n_1 - 2m_1 - m_t)! (n_2 - 2m_2 - m_t)!$. Thus,
\be \label{bipoint}
\begin{split}
\langle \zeta^{n_1}(\x_1) \zeta^{n_2}(\x_2) \rangle = \sum_{m_1, m_2, m_t} & \frac{n_1! \, n_2! \left(\sigma_{\zeta}^2|_{\x_1}\right)^{m_1} \left(\sigma_{\zeta}^2(x) \right)^{m_t} \left(\sigma_{\zeta}^2|_{\x_2}\right)^{m_2}  }{2^{m_1} m_1! \, (n_1 - 2m_1 - m_t)! \, m_t! \, (n_2 - 2m_2 - m_t)! \,  2^{m_2} m_2!} \\ & \times \langle \zeta^{n_1 - 2m_1 - m_t}(\x_1) \zeta^{n_2 - 2m_2 - m_t}(\x_2) \rangle_c.
\end{split}
\ee
Careful inspection of this result reveals that~\eqref{bipoint} leads to a 2-point distribution analogous to what was obtained in~\cite{Chen:2018brw}, but with two points defining the filtering instead of one:
\be
\begin{split} \label{PDF-2-app}
\PDF (\zeta_1,\zeta_2,r) = \PDF_{\rm G}&(\zeta_1,\zeta_2,r)   \left.\Bigg[ 1 + \int_\x \! \int_{-\infty}^{\infty} \!\!\!\! \de \bar \zeta   \frac{\exp \Big[{-\frac{ \left(\bar \zeta - \zeta(r,r_1,r_2) \right)^2}{2\sigma_\zeta^2 (r,r_1,r_2)} } \Big] }{\sqrt{2\pi  } \sigma_\zeta (r,r_1,r_2)} \right. \times \\ & \left. \left\{  \frac{W(r_1)}{s^2(r_1)} \!  \left( G_{11} \frac{\partial}{\partial {\bar \zeta}} - G_{12} \right)  +  \frac{W(r_2)}{s^2(r_2)} \!  \left( G_{21} \frac{\partial}{\partial {\bar \zeta}} - G_{22} \right) \right\} F  \! \left({\bar \zeta}\right)  \right.\Bigg],
\end{split}
\ee
where the coefficients $G_{ij}$, defined right below, depend on both the field variables and the spacetime positions $\x, \x_1, \x_2$ via the scalar variables
\be\nn
r\equiv|\x_1 - \x_2|, \quad r_1\equiv|\x - \x_1|,\quad r_2\equiv|\x - \x_2|.
\ee

In obtaining this expression, we have defined a number of functions that depend uniquely on the structure of the Gaussian theory. The variance $\sigma_{\zeta}^2(r,r_1,r_2)$ and $\zeta(r,r_1,r_2)$ are the regression coefficients obtained by conditioning a Gaussian distribution of $(\bar \zeta, \zeta_1, \zeta_2)$ over $(\zeta_1^W, \zeta_2^W)$, with covariance matrix given by~\eqref{cov}:
\begin{align}
\sigma_{W}^2(r,r_1,r_2) =& \, \sigma_{\zeta}^2 + \frac{  2 s^2(r_1) s^2(r_2)  }{\sigma_{W}^4 - \sigma_{W,{\rm ext}}^4(r)} \sigma_{W,{\rm ext}}^2(r)  - \frac{s^4(r_1)  + s^4(r_2)   }{\sigma_{W}^4 - \sigma_{W,{\rm ext}}^4(r)} \sigma_{W}^2, \\
\zeta(r,r_1,r_2) =& \, \frac{s^2(r_1)  \zeta_1  + s^2(r_2)  \zeta_2 }{\sigma_{W}^4 - \sigma_{W,{\rm ext}}^4(r)} \sigma_{W}^2  - \frac{ s^2(r_1) \zeta_2   +  s^2(r_2)  \zeta_1 }{\sigma_{W}^4 - \sigma_{W,{\rm ext}}^4(r)} \sigma_{W,{\rm ext}}^2(r).
\end{align}
Furthermore, the coefficients $G_{ij}$ are given by
\begin{align}
G_{11} &\equiv \sigma_\zeta^2 - s^2(r_2) \frac{s^2(r_2) \sigma_W^2 - s^2(r_1)\sigma_{W,{\rm ext}}^2(r)}{\sigma_{W}^4 - \sigma_{W,{\rm ext}}^4(r)}, \\
G_{12} &\equiv {\bar \zeta} - s^2(r_2) \frac{ \sigma_W^2  \zeta_2 - \sigma_{W,{\rm ext}}^2(r) \zeta_1}{\sigma_{W}^4 - \sigma_{W,{\rm ext}}^4(r)}, \\
G_{21} &\equiv \sigma_\zeta^2 - s^2(r_1) \frac{s^2(r_1) \sigma_W^2 - s^2(r_2)\sigma_{W,{\rm ext}}^2(r)}{\sigma_{W}^4 - \sigma_{W,{\rm ext}}^4(r)}, \\
G_{22} &\equiv {\bar \zeta} - s^2(r_1) \frac{ \sigma_W^2  \zeta_1 - \sigma_{W,{\rm ext}}^2(r) \zeta_2 }{\sigma_{W}^4 - \sigma_{W,{\rm ext}}^4(r)}.
\end{align}

\subsection{2-Point PDF: map to the sphere $S^2$}

The only difference with the previous Appendix is that the window function should be a map onto an angular coordinate $\n$ instead of a three-dimensional (flat) space. In the text, we wrote
\be
\begin{split}
\PDF_\Theta (\Theta_1,\Theta_2,\n_1,&\n_2) = \PDF_{G,W}(\Theta_1,\Theta_2,\n_1,\n_2)   \Bigg[ 1 - \int_\x \int_{-\infty}^{\infty} \!\!\!\! \de \bar \zeta   \frac{\exp \Big[{-\frac{ \left(\bar \zeta - \zeta_\Theta(\x,\n_1,\n_2) \right)^2}{2\sigma_\Theta^2 (\x,\n_1,\n_2)} } \Big] }{\sqrt{2\pi  } \sigma_\Theta (\x,\n_1,\n_2)} \;\times \\ &   \left\{  \frac{W_\Theta(\x, \n_1)}{s_\Theta^2(\x , \n_1)} \!  \left( G_{11}^{\Theta} \frac{\partial}{\partial {\bar \zeta}} - G_{12}^{\Theta} \right)      +  \frac{W_\Theta(\x, \n_2)}{s_\Theta^2(\x,  \n_2)} \!  \left( G_{21}^{\Theta} \frac{\partial}{\partial {\bar \zeta}} - G_{22}^{\Theta} \right) \right\} F  \! \left({\bar \zeta}\right)  \Bigg].
\end{split}
\ee
Here we have
\begin{align}
W_\Theta(\x,\n) \equiv \int_k e^{i \K \cdot \x} T(k,\n) &\qquad\text{and} & s^2_\Theta(\x,\n) \equiv \int_k e^{i \K \cdot \x} T(k,\n) P_\zeta(k),
\end{align}
while the regression coefficients are given by
\begin{align}
\sigma_{\Theta}^2(\x,\n_1,\n_2) =& \, \sigma_{\zeta}^2 + \frac{  2 s_\Theta^2(\x,\n_1) s_\Theta^2(\x,\n_2)  }{\sigma_{\Theta}^4 - \sigma_{\Theta,{\rm ext}}^4(\n_1,\n_2)} \sigma_{\Theta,{\rm ext}}^2(\n_1,\n_2) \nn \\ & - \frac{s^4(\x,\n_1)  + s^4(\x,\n_2)   }{\sigma_{\Theta}^4 - \sigma_{\Theta,{\rm ext}}^4(\n_1,\n_2)} \sigma_{\Theta}^2, \\
\zeta_\Theta(\x,\n_1,\n_2) =& \, \frac{s_\Theta^2(\x,\n_1)  \Theta_1  + s_\Theta^2(\x,\n_2)  \Theta_2 }{\sigma_{\Theta}^4 - \sigma_{\Theta,{\rm ext}}^4(\n_1,\n_2)} \sigma_{\Theta}^2 \nn \\ & - \frac{ s_\Theta^2(\x,\n_1) \Theta_2   +  s_\Theta^2(\x,\n_2)  \Theta_1 }{\sigma_{\Theta}^4 - \sigma_{\Theta,{\rm ext}}^4(\n_1,\n_2)} \sigma_{\Theta,{\rm ext}}^2(\n_1,\n_2).
\end{align}
Finally, the functions $G_{ij}^\Theta$ read
\begin{align}
G_{11}^\Theta &\equiv \sigma_\zeta^2 - s_\Theta^2(\x,\n_2) \frac{s_\Theta^2(\x,\n_2) \sigma_\Theta^2 - s_\Theta^2(\x,\n_1)\sigma_{\Theta,{\rm ext}}^2(\n_1,\n_2)}{\sigma_{\Theta}^4 - \sigma_{\Theta,{\rm ext}}^4(\n_1,\n_2)}, \\
G_{12}^\Theta &\equiv {\bar \zeta} - s_\Theta^2(\x,\n_2) \frac{ \sigma_\Theta^2  \Theta_2 - \sigma_{\Theta,{\rm ext}}^2(\n_1,\n_2) \Theta_1}{\sigma_{\Theta}^4 - \sigma_{\Theta,{\rm ext}}^4(\n_1,\n_2)}, \\
G_{21}^\Theta &\equiv \sigma_\zeta^2 - s_\Theta^2(\x,\n_1) \frac{s_\Theta^2(\x,\n_1) \sigma_\Theta^2 - s_\Theta^2(\x,\n_2)\sigma_{\Theta,{\rm ext}}^2(\n_1,\n_2)}{\sigma_{\Theta}^4 - \sigma_{\Theta,{\rm ext}}^4(\n_1,\n_2)}, \\
G_{22}^\Theta &\equiv {\bar \zeta} - s_\Theta^2(\x,\n_1) \frac{ \sigma_\Theta^2  \Theta_1 - \sigma_{\Theta,{\rm ext}}^2(\n_1,\n_2) \Theta_2 }{\sigma_{\Theta}^4 - \sigma_{\Theta,{\rm ext}}^4(\n_1,\n_2)}.
\end{align}

\end{appendix}

\end{document}